\documentclass[12pt]{article}
\usepackage{a4wide}
\usepackage{graphicx}
\usepackage{amssymb}
\usepackage{hyperref}

\newcommand{\be}{\begin{equation}}
\newcommand{\ee}{\end{equation}}
\newcommand{\ba}{\begin{eqnarray}}
\newcommand{\ea}{\end{eqnarray}}

\begin{document}

\begin{titlepage}
\begin{flushright}
LU TP 11-14\\
arXiv:1103.5945 [hep-ph]\\
Revised September 2011
\end{flushright}
\vfill
\begin{center}
{\Large\bf A new global fit of the $L^r_i$ at next-to-next-to-leading order in
Chiral Perturbation Theory}
\vfill
{\bf Johan Bijnens and Ilaria Jemos}\\[0.3cm]
{Department of Astronomy and Theoretical Physics, Lund University,\\
S\"olvegatan 14A, SE 223-62 Lund, Sweden}
\end{center}
\vfill
\begin{abstract}
A new fit is done to obtain numerical values for the order $p^4$ low-energy-constants $L_i^r$
in Chiral Perturbation Theory. This includes both new data and new calculated observables.
We take into account masses, decay constants, $K_{\ell4}$,
$\pi\pi$ and $\pi K$ scattering lengths and slopes and the slope of the pion scalar formfactor.
We compare in detail where the changes w.r.t. to the 10 year old ``fit 10'' come from.
We discuss several scenarios for estimating the order $p^6$ constants $C_i^r$ and search for
possible values of them that provide a good convergence for the ChPT series.
We present two such sets. One big change is that the fits do not have the expected behaviour
in the limit of large $N_c$ as well as before.
\end{abstract}
\vfill
\vfill
\end{titlepage}

\section{Introduction}

Since its very beginning Chiral Perturbation Theory (ChPT)\cite{Weinberg0,GL1,GL2}, the
  effective field theory of QCD at low energies, has been successful in
  the description of several hadronic observables. Unfortunately when one
  tries to perform loop calculations to improve the precision of the
  predictions, one faces a problem. The couplings appearing in the
  $\mathcal{L}_4$ Lagrangian are many, i.e. 10, and they must be determined from
  phenomenology. One of the first determinations of such couplings was
  done already in \cite{GL2}. There most of the next-to-leading order (NLO)
  couplings were inferred both from
  phenomenology and from considerations lead by large $N_c$ estimates (where $N_c$
  is the  number of colours).

Considering such good results it is important to decide
  whether ChPT is a suitable theory to achieve precise determinations of the
  hadronic observables. It urged then to carry
  on a program and perform next-to-next-to-leading order (NNLO)
  calculations \cite{Bijnens:1999sh}. In the last 10 years many two-loop calculations in
  three-flavour ChPT have been done, see \cite{reviewp6} for a review.

Notice however that going to higher orders raises a serious issue: the number of
  the unknown couplings increases rapidly. If on the one side adding loop
  diagrams should allow us to include better corrections and improve our
  descriptions, on the other side, many unknown parameters contribute and this
  seriously threatens the predictivity of the theory. Furthermore without
  knowing the values of such constants the convergence of the chiral
  expansions are difficult to test, although feasible with the method
  described in \cite{Bijnens:2009zd}.

The two-loop expressions now available
can be used to perform a new global fit at NNLO of some of the 
next-to-leading-order (NLO) low-energy-constants (LECs) into the game, 
the $L^r_i$. A first attempt was done
\cite{Bijnens:1994ie} when some experimental information was available for the $K_{\ell4}$
decay and by estimating the NNLO contributions using dispersive
analysis. The fit  was refined later on, when the full NNLO
calculation for this process was performed
\cite{Amoros:2000mc,Amoros:2001cp}. 
After that many other observables have
been calculated at NNLO and many of them are also better known
experimentally. Therefore the time has come to perform a new fit of the
$L^r_i$ couplings at a NNLO precision. In this paper we present results for such
a fit. Some studies using the extra observables but without performing a full new fit
were reported in \cite{Bijnens:2003xg,BDT,BDT2}. Notice that in a preliminary
phase we converted all the FORTRAN programs used to evaluate
the amplitudes up to NNLO into C++ code. Our fits are all performed using such programs.

The paper is organized as follows. In Section~\ref{sect:ChPT} we sketch out
the ChPT formalism and its main underlying ideas. In Section~\ref{sect:input} we
 review the phenomenological input we used in our results. We also show
 which are the $L^r_i$ that give the largest contributions  for each observable
 we included. Notice that now much more input is present compared to the past
 fits \cite{Amoros:2000mc,Amoros:2001cp}. In Section~\ref{sect:Cires} we
 present the main model we used to estimate the $C^r_i$, i.e. the coupling
 constants appearing in the NNLO Lagrangian. Such an estimate is usually
 called the resonance estimate. In Section~\ref{sect:fit10} we summarize the status
 of the main existing NNLO fit so far: fit 10 of \cite{Amoros:2001cp}. In
 Section~\ref{sect:newfits} we show our main findings using the $C^r_i$
 estimates of Section~\ref{sect:Cires}. We quote different fits
 so to show how each observable we include affects our findings. The best fit
 we get we call fit All and should be considered the main output of this work.
 This fit exhibits several differences with fit
 10. One especially striking feature is that it does not respect any longer
 the large $N_c$ relation $2L^r_1\approx L^r_2$. We also
 show for fit All the convergence of the expansions for masses and decay
 constants, which is much improved compared to the one of fit 10. In
 Section~\ref{sect:linfit} we perform a fit of the $L^r_i$ using as input
 different experimental results for the $K_{\ell4}$ amplitude. We show that
 with this input the  large $N_c$ relation $2L^r_1\approx L^r_2$ is better
 satisfied, although the resulting fit is not as good as fit All in convergence for the masses.
 In Section~\ref{sect:confres} we try to justify and test our estimate of the $C^r_i$. 
 In Section~\ref{sect:li} we compare
 further the fits obtained using their predictions for the two-flavour LECs
 $\bar{\ell}_i$. Fit All again results as the most convincing one. In
 Section~\ref{sect:chinaCi} we show results for fits based on a different
 estimate of the $C^r_i$ couplings that can be found in \cite{Jiang:2009uf}, this is essentially
  a chiral quark model estimate (CQM). The
 fits are not as good as fit All, nor for the $\chi^2$ nor for the
convergence of the expansions. Also in this case the large $N_c$ relation is
not satisfied. In Section~\ref{sect:comparison} we provide a short
comparison with the recent determination of $L_5^r$ of \cite{Ecker:2010nc}.

In Section~\ref{sect:randomCi} we show results for an another
treatment of the $C^r_i$. We let the $C^r_i$ couplings to take random values,
although they are forced to keep the size $1/(16\pi^2)^2$. These fits have been done requiring
 extra constraints of convergence for mass and decay constant expansions, as explained 
in Section~\ref{sect:conv}. In this way it is easier
to select only credible fits. The results of such a study are finally
shown in Section~\ref{sect:randres}. We found very many good fits that
correctly predict all the observables used as input and with low
$\chi^2$. These fits are unfortunately different looking from each other. Therefore we can
only show which are the ranges where we found the $L^r_i$ to vary. For some of
the NLO constants such ranges are very wide. This method
shows however that it is possible to fit the NNLO expressions to the
observables with $C^r_i$ of the expected size and it also allows to study well the
strong correlations between the couplings.  
Finally in the appendix we present a table where we quote our estimates
for the NNLO couplings.

\section{Chiral Perturbation Theory}
\label{sect:ChPT}

We devote this section to a brief description of the formalism of
three-flavour ChPT \cite{Weinberg0,GL1,GL2}. 
Introductory references are \cite{Pich,Scherer1}.
The notation in the following is the same as in \cite{Bijnens:1999sh}.
ChPT relies on the assumption that the flavour symmetry of QCD is
spontaneously broken to the diagonal subgroup, $SU(3)_L\times
SU(3)_R\rightarrow SU(3)_V$. According to the Goldstone theorem, 
$8$ pseudo Goldstone bosons then
arise. These are identified with the low lying pseudoscalar mesons and are
organized in a unitary  $3\times 3$ matrix
\be
u =\exp\left(\frac{i}{\sqrt{2}F} \phi \right)\,,
\ee
where $\phi$ is a hermitian $3\times 3$ matrix:
\ba
\phi =
 \left( \begin{array}{ccc}
\frac{1}{\sqrt{2}}\pi^0  + \frac{1}{\sqrt{6}} \eta &  \pi^+ & K^+ \\
 \pi^- & -\frac{1}{\sqrt{2}}\pi^0 + \frac{1}{\sqrt{6}} \eta  & K^0 \\
 K^- & \bar{K}^0 & -\frac{2}{\sqrt{6}} \eta  \\
 \end{array} \right)\,.
\ea
The Lagrangian describing the low-momentum strong interactions of the light
mesons  must be invariant under $SU(3)_L\times
SU(3)_R$ local transformations. The most general lowest order Lagrangian is  
\be
\label{pilagrangian}
\mathcal{L}_{2}= \frac{F^2_0}{4}  
\left( \langle u_{\mu} u^{\mu} \rangle   + \langle \chi_{+}\rangle \right),
\ee
with
\ba
 u_{\mu} &=& i\{  u^{\dag}( \partial_{\mu} - i r_{\mu}   )u
 -u ( \partial_{\mu}   -i l_{\mu}    ) u^{\dag}   \}\,,
\nonumber\\
\chi_{\pm} &=& u^{\dag} \chi u^{\dag} \pm u \chi^{\dag} u\,,
\nonumber\\
 \chi &=& 2B_0 (s+ip)\,.\nonumber
\ea
The fields $s$, $p$, $l_\mu=v_\mu-a_\mu$ and
$r_\mu=v_\mu+a_\mu$ are the standard external scalar, pseudoscalar, left- and
right-handed vector fields introduced by Gasser and Leutwyler \cite{GL1,GL2}.
The constants $F_0$ and $B_0$ are instead the leading-order (LO) LECs. The notation $\langle
X\rangle$ stands for trace over up, down and strange quark flavour.

Starting from this Lagrangian we can then build up an effective field theory
by including loop diagrams and higher order Lagrangians, where operators of
higher dimensions are included. Their coupling constants are then counter-terms,
i.e. their infinities absorb the UV divergences coming
from the loop diagrams. In this way one obtains a theory renormalized
order by order.
Unfortunately going to higher orders the number of operators allowed by the
symmetries increase and therefore also the number of unknown coupling
constants. At NLO there are 10 LECs, called $L^r_i$, 
while at NNLO there are as many as 94, called $C^r_i$.
We will always quote the renormalized versions where the $C^r_i$ are made
dimensionless by using the physical value of $F_\pi$.
The renormalization scale is chosen to be $\mu=770~$MeV.
While there is in principle enough phenomenological information to fit the
first ones, we still need to rely on theoretical models or on some other
method for the latter ones, as those described in
Sections~\ref{sect:Cires},~\ref{sect:chinaCi} and \ref{sect:randomCi}.   

\section{Fitting procedure and input observables}
\label{sect:input}

In this section we first show how we perform the fits and then we review
shortly the observables we use as input and their values.

\subsection{$\chi^2$}

The fit is performed using MINUIT in its C++ version
\cite{James:1975dr,MINUIT}. MINUIT is a routine to find the minimum value of a
multi-parameter function. The procedure to perform the fits is very similar to
the one explained in \cite{Amoros:2001cp}. The
function to be minimized is the $\chi^2$ of the fit. It is obtained
summing up the partial contributions $\chi^2_{i({\rm part})}$ for each input
observables
\ba
\chi^2_{i ({\rm part})}&=&\left(\frac{x_{i ({\rm meas})}-x_{i ({\rm calc})}}
 {\Delta x_i}\right)^2
\nonumber\\
\chi^2&=&\sum_i \chi^2_{i ({\rm part})}
\ea
where $x_{i ({\rm meas})}$ are the physical values for each input observables
  and $\Delta x_i$ their associated errors. $x_{i ({\rm calc})}$ are the
    results as calculated by ChPT up to NNLO. The rest of this section is
    devoted to list the
    values and the uncertainties used for each $x_i$.

\subsection{Masses and decay constants}
\label{sect:massesanddecay}

The masses and decay constants of the light pseudoscalar mesons have been
calculated at NNLO in \cite{Amoros:1999dp}. 
We use them as physical parameters, namely  as input to
calculate the several observables, similarly to what was done in
\cite{Amoros:2001cp}. Their values are given in \cite{Nakamura:2010zzi} and are
\ba
\label{masses}
m_{\pi^+}&=&139.57018\,\,{\rm MeV},\quad m_{\pi^0}=134.9766\,\,{\rm MeV},
  \quad m_{\eta}=547.853\,\,{\rm MeV},\nonumber\\ 
m_{K^+}&=&493.677\,\,{\rm MeV},\quad m_{K^0}=497.614\,\,{\rm MeV},
\\\label{decayconst}
F_\pi&=&0.0922\pm0.0002\,\,{\rm GeV}\,.
\ea
The measurements in (\ref{masses}) and (\ref{decayconst}) 
differ slightly from the ones used in the latest full fit \cite{Amoros:2001cp}.
\ba
\label{massesold}
m_{\pi^+}&=&139.56995\,\,{\rm MeV},\quad m_{\pi^0}=134.9764\,\,{\rm MeV},
\quad m_{\eta}=547.30\,\,{\rm MeV},\nonumber\\ 
m_{K^+}&=&493.677\,\,{\rm MeV},\quad m_{K^0}=497.672\,\,{\rm MeV},
\\
\label{decayconstold}
F_\pi&=&0.0924\,\,{\rm GeV}\,.
\ea
Notice that preliminary results of our work have been reported in the
proceedings \cite{Bijnens:2009hy,Bijnens:2009pm} and several unpublished talks. 
Those results were based on the masses of \cite{Amsler:2008zzb}, 
that differ slightly from both
(\ref{masses}), (\ref{decayconst}) and (\ref{massesold}), (\ref{decayconstold}).

All the new fits shown in this paper have been produced using the
values in (\ref{masses}) and (\ref{decayconst}). However we have not observed
any substantial modification of the outputs when using the old masses
 (\ref{massesold})
and pion decay constant (\ref{decayconstold}). As discussed below other changes
in experimental input are behind the changes of central values.

The masses depend at LO on $B_0\hat{m}$ (with $\hat{m}=(m_u+m_d)/2$) and on
$B_0 m_s$. $L^{r}_4$, $L^{r}_6$, $L^{r}_5$ and $L^{r}_8$ appear in the
 expression
at NLO. In $m^2_\eta$ there is also a NLO contribution from $L^{r}_7$. 
The decay constant
$F_\pi$ depends instead on $F_0$, as an overall factor,  and on
$L^{r}_4$ and $L^{r}_5$.

\subsection{$F_K/F_\pi$}
\label{sect:fkfpi} 

As input observable for our fits we will use the ratio $F_K/F_\pi$ to
eliminate the dependence on the unknown constant $F_0$, since
 it contributes as an
overall factor for $F_K$ as well. In the end the value of $F_\pi$ 
then determines the value for $F_0$.

The ratio takes the value \cite{Nakamura:2010zzi}
\be
\label{ratio}
\frac{F_K}{F_\pi}=1.197 \pm 0.007,
\ee
that is also in full agreement with several lattice estimates as reported
in \cite{Colangelo:2010et}. Using $F_K/F_\pi$ at NLO we are sensitive to $L^r_5$. To perform the fit we expand the ratio as
\be
\label{ratio1}
\frac{F_K}{F_\pi}=1+
\underbrace{
\left.\frac{F_K}{F_0}\right|_{p^4}-\left.\frac{F_{\pi}}{F_0}\right|_{p^4}}
_{\rm NLO}
\underbrace{
+\left.\frac{F_K}{F_0}\right|_{p^6}-\left.\frac{F_\pi}{F_0}\right|_{p^6}
-\left.\frac{F_K}{F_0}\right|_{p^4} \left.\frac{F_{\pi}}{F_0}\right|_{p^4}
+\left.\frac{F_\pi}{F_0}\right|^2_{p^4}}
_{\rm NNLO}\,,
\ee
so that we can keep track of the exact contributions from the different
orders. However we also check that the same quantity estimated as
\be\label{ratio2}
\frac{F_K}{F_\pi}=\frac{F_0+\left.F_K\right|_{p^4}+\left.F_K\right|_{p^6}}
{F_0+\left.F_\pi\right|_{p^4}+\left.F_\pi\right|_{p^6}}
\ee
gives approximately the same value\footnote{We thank Veronique
Bernard and Emilie Passemar for pointing out that these were significantly
different for some of our preliminary fits.}.

Notice the experimental result for the ratio $F_K/F_\pi$
 (\ref{ratio}) differs substantially from the one
used in \cite{Amoros:2001cp}. Ref.~\cite{Amoros:2001cp} used
$F_K/F_\pi=1.22\pm0.01$. The change in $F_K/F_\pi$ is one
of the major sources of difference with \cite{Amoros:2001cp} 
as will be shown later in Section~\ref{sect:newfits}.

\subsection{The quark-mass ratio $m_s/\hat{m}$}

For the masses we have a similar problem, they depend on the quark masses
and on $B_0$.  We thus use
as was done in \cite{Amoros:2000mc,Amoros:2001cp}
the ratio of the strange quark mass over the isospin doublet quark mass
$\hat{m}$ as input observable.  
The two following relations
involving the light pseudoscalar meson masses hold at LO in ChPT
\be \label{massratio}
\left.\frac{m_s}{\hat{m}}\right|_1=\frac{2m^2_{0K}-m^2_{0\pi}}{m^2_{0\pi}}
\qquad
\left.\frac{m_s}{\hat{m}}\right|_2=\frac{3m^2_{0\eta}-m^2_{0\pi}}{2m^2_{0\pi}}
\ee
where with $m_0$ we indicate the meson masses at LO. They are calculated
subtracting from the physical values the NLO and NNLO expressions. We include
both relations in (\ref{massratio}) in the fits. 
For the pion mass we use the neutral pion mass
$m_{\pi^0}$. In the kaon case we need to correct the physical value
for the mass since its electromagnetic contribution is sizeable. 
We take the average between $m_{K^+}$ and
$m_{K^0}$ and then we subtract the electromagnetic contribution as stated by
the Dashen's theorem and an estimate of its violation:
\be\label{mkelm}
m^2_{K{\rm
    av}}=\frac{1}{2}(m^2_{K^+}+m^2_{K^0}-1.8(m^2_{\pi^+}-m^2_{\pi^0}))=(494.50
\,\,\rm{ MeV})^2.
\ee
The factor 1.8 in (\ref{mkelm}) is due to the corrections to Dashen's theorem
where we use the value of \cite{Bijnens:1996kk}.

The value of the quark mass ratio has been calculated by several lattice
collaborations. The authors of \cite{Amoros:2001cp} used as standard input
$m_s/\hat{m}=24$ with a $10\%$ uncertainty, but they also checked that
$m_s/\hat{m}=26$ was compatible. Here instead we use $m_s/\hat{m}=27.8$ as
obtained by the Flavianet Lattice Averaging Group in \cite{Colangelo:2010et},
and we again adopt a $10\%$ uncertainty for the error to be used in the fits
when comparing with the theoretical values (\ref{massratio}).

In the end the calculated NLO and NNLO masses are used to determine
the lowest order
mass or alternatively $B_0\hat m$.
 
\subsection{$K_{\ell 4}$ formfactors}
\label{sect:kl4}

The decay $K^+(p)\to \pi^+(p_1)\pi^-(p_2) e^+(p_\ell)\nu(p_\nu)$ is given 
by the amplitude~\cite{Bijnens:1994ie}
\be
      T = \frac{G_F}{\sqrt{2}} V^\star_{us} \bar{u} (p_\nu) \gamma_\mu
      (1-\gamma_5) v(p_\ell) (V^\mu - A^\mu).
      \label{k11}
\ee
In (\ref{k11}) $V^\mu$ and $A^\mu$ can be parametrized in terms of four
formfactors: $F$,
$G$, $H$ and $R$. However the $R$-formfactor is negligible in decays with
an electron in the final state.
Using partial wave expansion and neglecting $d$ wave terms one
obtains for the $F$, $G$ and the $H$ formfactors \cite{Amoros:1999mg}:
\ba
\label{Kl4exp}
F(s_{\pi},s_{\ell},\cos\theta)&=&f_{s}(s_\pi,s_{\ell})e^{i\delta_s}
+f_{p}e^{i\delta_p}\cos\theta+\ldots
\,,
\nonumber\\
G(s_{\pi},s_{\ell},\cos\theta)&=&g_{p}(s_\pi,s_{\ell})e^{i\delta_p}+\ldots\, ,
\nonumber\\
H(s_{\pi},s_{\ell},\cos\theta)&=&h_{p}(s_\pi,s_{\ell})e^{i\delta_p}+\ldots\, ,
\ea
where we also assumed that the $p$ phase is the same for the three formfactors.
In (\ref{Kl4exp}) $s_{\pi}(s_{\ell})$ is the invariant
mass of dipion (dilepton) system,
 $\theta$ is the angle of the pion in their rest frame
w.r.t. the kaon momentum. The $F$ and $G$ formfactors were calculated at NNLO in
\cite{Amoros:2000mc}. The quantities are especially sensitive to $L^r_1$,
$L^r_2$ and $L^r_3$. Also the $H$ formfactor
 is known at order $p^6$ \cite{ABBC}
but we do not use it as input observable since 
it depends on a different set of LECs,
those from the anomalous intrinsic parity sector.

The measured observables are obtained by further parametrizing
$f_s(s_\pi,s_{\ell})$ and $g_p(s_\pi,s_{\ell})$ as
\ba\label{kl4par}
f_s(s_\pi,s_{\ell})&=&f_s+f'_{s}q^{2}+f''_{s}q^{4}+f'_es_{\ell}/4m^{2}_{\pi},
\nonumber\\
g_p(s_\pi,s_{\ell})&=&g_{p}+g'_{p}q^{2},
\ea
where $ q^{2}=s_{\pi}/(4m^{2}_{\pi})-1$. 
(\ref{kl4par}) can be used to fit the
measured data points. In \cite{Amoros:2001cp} the preliminary linear fit
from the E865 measurement \cite{Truoel} was used as input. It has the values
\be\label{kl4oldmeas}
f_s=5.77\pm0.097,\quad f'_s=0.47\pm0.15,
\quad g_p=4.684\pm0.092,\quad g'_p=0.54\pm0.20. 
\ee
Now more precise results from the NA48/2 experiment are available
\cite{Batley:2010zza} and their second order fit of the formfactors read
\be\label{kl4newmeasratio}
\frac{f'_s}{f_s}=0.152\pm0.009,\quad \frac{f''_s}{f_s}=-0.073\pm0.009,
\quad\frac{g_p}{f_s}=0.868\pm0.01,\quad\frac{g'_p}{f_s}=0.089\pm0.02,
\ee
Notice that in \cite{Batley:2010zza} no  measure of  $f_s$ is reported, 
therefore  we always use the $f_s=5.75\pm0.097$ from the E865
collaboration \cite{Pislak:2001bf} in our fits. Multiplying by $f_{s}$
 and combining the errors in quadrature we obtain as measures 
\ba
\label{kl4newmeas}
f_s&=&5.750\pm0.097,\quad
f'_s=0.874\pm0.054,\quad 
f''_s=-0.420\pm0.052,
\nonumber\\
g_p&=&4.99\pm0.12,\quad
g'_p=0.512\pm0.121\,.
\ea

In \cite{Batley:2010zza} there is also a linear fit which gives
$\frac{f'_s}{f_s}=0.073\pm0.004$ and thus $f'_{s}=0.420\pm0.024$. 
Deciding which of the two fits for the $F_s$ formfactor should be
used is a relevant issue.
The problem is how much we can rely on the curvature of the formfactor
$F_s$. As a matter of fact it is difficult for NNLO ChPT to
reproduce the large negative curvature $f''_s$, as was also noted
in \cite{Amoros:2001cp,Bijnens:2009zd}. A dispersive analysis
approach combined to two loops ChPT, similar to the one done for $\pi\pi$
scattering \cite{CGL}, might clarify the situation. An indication of
 this is given by Figure~7 of \cite{Amoros:2001cp}. It is visible there
 that the dispersive result
for $K_{\ell 4}$ decay \cite{Bijnens:1994ie} has a larger curvature
 than the two-loop
result \cite{Amoros:2001cp}.

Let us conclude with a cautionary remark about the $K_{\ell 4}$ data. In
\cite{Colangelo:2008sm} it was made clear that isospin breaking effects at
threshold give important corrections. These have not been taken into
account in the NA48/2 analysis \cite{Batley:2010zza}, thus they might 
affect significantly their findings.

\subsection{$\pi\pi$ scattering}

The $\pi\pi$ scattering amplitude can be written as a function $A(s,t,u)$
which is symmetric in $t,\,u$:
\begin{equation}
A(\pi^{a}\pi^{b}\rightarrow\pi^{c}\pi^{d})=
\delta^{a,b}\delta^{c,d}A(s,t,u)+\delta^{a,c}\delta^{b,d}A(t,u,s)
+\delta^{a,d}\delta^{b,c}A(u,t,s)\,,
\end{equation}
where $s,t,u$ are the usual Mandelstam variables.
The three flavour ChPT calculation of $A(s,t,u)$ was done in \cite{BDT}.
The isospin amplitudes $T^{I}(s,t)$ $(I=0,1,2)$ are
\ba
T^{0}(s,t)&=&3A(s,t,u)+A(t,u,s)+A(u,s,t)\,,\,\nonumber\\
T^{1}(s,t)&=&A(s,t,u)-A(u,s,t)\,,\,\nonumber\\
T^{2}(s,t)&=&A(t,u,s)+A(u,s,t)\,,\,
\ea
and are expanded in partial waves 
\ba
\label{partialwavepipi}
T^{I}(s,t)&=&32\pi\sum_{
  \ell=0}^{+\infty}(2\ell+1)P_{\ell}(\cos{\theta})t_{\ell}^{I}(s),
\ea
where $t$ and $u$ have been written as 
$t=-\frac{1}{2}(s-4m^{2}_{\pi})(1-\cos{\theta})$,
 $u=-\frac{1}{2}(s-4m^{2}_{\pi})(1+\cos{\theta})$. In (\ref{partialwavepipi})
we indicate with  $P_{\ell}(\cos{\theta})$ the Legendre polynomials.
Near threshold the $t^I_{\ell}$ are further expanded in terms of the
 threshold parameters
\begin{eqnarray}
t_{\ell}^{I}(s)=q^{2\ell}( a_{\ell}^{I}
+ b_{\ell}^{I}q^{2}+\mathcal{O}(q^{4})),
\qquad
 q^{2}=\frac{1}{4}(s-4m^{2}_{\pi}) ,
\end{eqnarray}
where $a_{\ell}^{I}, b_{ \ell}^{I}\dots$ are the 
scattering lengths, slopes,$\dots$. These thresholds parameters constitute our
observables.
Currently a very precise determination of these parameters exists. It is based
on a dispersive analysis approach and on two-flavour ChPT and can be found in
\cite{CGL}. In Table~\ref{tab:pipi} we quote the values of the threshold
parameters we use in our fits and their corresponding uncertainties, 
which we took to be double the ones in \cite{CGL}.
For most fits we used only $a^0_0$ and $a^2_0$ but we have checked that the
others listed in Table~\ref{tab:pipi} are also well within the uncertainties
quoted.
\begin{table}
\begin{center}
\begin{tabular}{|c|c|c|}
\hline
\rule{0cm}{15pt}
$a^0_0$ & $0.220  \pm 0.010$ & $m^0_\pi$          \\[2pt]
\hline
\rule{0cm}{15pt}
$b^0_0$ & $0.276  \pm 0.012$ & $m^{-2}_\pi$        \\[2pt]  
\hline
\rule{0cm}{15pt}
$a^2_0$ & $-0.444 \pm 0.020$ & $10^{-1} m^{0}_\pi$  \\[2pt]
\hline
\rule{0cm}{15pt}
$b^2_0$ & $-0.803 \pm 0.024$ & $10^{-1} m^{-2}_\pi$ \\[2pt]
\hline
\rule{0cm}{15pt}
$a^1_1$ & $0.379  \pm 0.010$ & $10^{-1}m^{-2}_\pi$  \\[2pt]
\hline
\rule{0cm}{15pt}
$b^1_1$ & $0.567  \pm 0.026$ & $10^{-2}m^{-4}_\pi$  \\[2pt]
\hline
\end{tabular}
\caption{\label{tab:pipi} The values of the scattering lengths and slopes as
found in \cite{CGL} and our fitting uncertainties. 
In the third column the normalization
factors are given. We quote here only those scattering parameters added as
 input in our fits.}
\end{center}
\end{table}
Notice also that the NA48/2 experiment in \cite{Batley:2010zza} obtained
compatible values for $a^0_0$ and
$a^2_0$ from the measurement of the $\delta=\delta_p-\delta_s$ phase shift in
$K_{\ell 4}$ decays.

\subsection{$\pi K$ scattering}

The $\pi K$ scattering process has amplitudes $T^I(s,t,u)$
in the isospin channels $I= 1/2,3/2$. They have been calculated at NNLO in
ChPT in \cite{BDT2}. As for $\pi \pi$ scattering, it is possible to define
scattering lengths and
 slopes $a_{\ell}^{I}$, $b_{\ell}^{I}$. So we introduce the partial
 wave expansion of
 the isospin amplitudes
\be
T^{I}(s,t,u) = 16\pi
\sum_{\ell=0}^{+\infty}(2\ell+1)P_{\ell}(\cos{\theta})t_{\ell}^{I}(s),
\ee
where $P_{\ell}(\cos{\theta})$ are the Legendre polynomials. Then  we expand
the $t_{\ell}^{I}(s)$ near threshold
\be\label{tilpiK}
t_{\ell}^{I}(s)=\frac{1}{2}\sqrt{s}q_{\pi K}^{2\ell}
\left(a_{\ell}^{I}+b^{I}_{\ell}q^{2}_{\pi K}+\mathcal{O}(q^{4}_{\pi K})\right),
\ee
where
\be\label{qpiK}
q_{\pi K}^{2}=\frac{s}{4}\left(1-\frac{(m_{K}+m_{\pi})^{2}}{s}\right)
          \left(1-\frac{(m_{K}-m_{\pi})^{2}}{s}\right)\,,
\ee
is the magnitude of the three-momentum in the center of mass system.
The Mandelstam variables are given in terms of the scattering angle $\theta$ by
\be
t=-2q^{2}_{\pi K}(1-\cos{\theta}),\quad u=-s-t+2m^{2}_{K}+2m^{2}_{\pi}\,.
\ee
(\ref{tilpiK}) defines the $\pi K$ scattering parameters $a^I_\ell$ and
$b^I_\ell$ that are our input observables. These have been computed from Roy
and Steiner type equations in \cite{BDM}. 
\begin{table}
\begin{center}
\begin{tabular}{|c|c|c|}
\hline
\rule{0cm}{15pt}
$a^{1/2}_0$ & $0.224  \pm 0.044$ & $m^{-1}_\pi$      \\[2pt]
\hline
\rule{0cm}{15pt}
$b^{1/2}_0$ & $0.85  \pm 0.08$ & $10^{-1}m^{-3}_\pi$ \\[2pt]  
\hline
\rule{0cm}{15pt}
$a^{3/2}_0$ & $-0.448 \pm 0.154$ & $10^{-1}m^{-1}_\pi$  \\[2pt]
\hline
\rule{0cm}{15pt}
$b^{3/2}_0$ & $-0.37 \pm 0.06$ & $10^{-1} m^{-3}_\pi$ \\[2pt]
\hline
\rule{0cm}{15pt}
$a^{1/2}_1$ & $0.19  \pm 0.02$ & $10^{-1}m^{-3}_\pi$  \\[2pt]
\hline
\rule{0cm}{15pt}
$b^{1/2}_1$ & $0.18  \pm 0.04$ & $10^{-2}m^{-5}_\pi$  \\[2pt]
\hline 
\rule{0cm}{15pt}
$a^{3/2}_1$ & $0.65  \pm 0.88$ & $10^{-3}m^{-3}_\pi$  \\ [2pt]
\hline
\rule{0cm}{15pt}
$b^{3/2}_1$ & $-0.92  \pm 0.34$ & $10^{-3}m^{-5}_\pi$  \\ [2pt]
\hline
\end{tabular}
\caption{\label{tab:piK}The values of the scattering lengths and slopes as
  found in \cite{BDM}. The uncertainties quoted here are those used
 in our fits and
  are double the ones of \cite{BDM}. In the third column the normalization
  factors are given. 
  We quote only those scattering parameters added as input in our fits.}
\end{center}
\end{table}
The results for the $s$ and $p$
waves scattering parameters we use are reported in Table~\ref{tab:piK}.
For most numerical results we used only $a^{1/2}_0$ and $a^{3/2}_0$
 but we have checked
that all the others agree within uncertainties.

\subsection{Scalar formfactor}
\label{scalarFF}

The scalar formfactor for the pion is defined as
\begin{equation}
F_{S}^{\pi}(t)=\langle \pi^0(p)|\bar{u}u+\bar{d}{d}|\pi^0(q)\rangle,
\end{equation}
where $t=p-q$. Near $t=0$ it is expanded via
\be
F^\pi_S(t) = F^\pi_S(0)\left(1+\frac{1}{6}\langle r^2 \rangle^\pi_S t
 + c^\pi_St^2+\ldots\right)\,.
\ee
The observables $\langle r^2 \rangle^\pi_S$ and $c^\pi_S$ are
used as input in our fits. The NNLO ChPT calculation for these
quantities was performed in \cite{Bijnens:2003xg}. The scalar
formfactor cannot be measured experimentally. Measuring the $\pi\pi$
phase shifts and using a dispersive representation it is possible to infer its
energy behaviour and therefore the
values of $\langle r^2 \rangle^\pi_S$
 and $c^\pi_S$ \cite{Donoghue:1990xh,Moussallam:1999aq,Ananthanarayan:2004xy}
\be\label{sff}
\langle r^2 \rangle^\pi_S=0.61\pm0.04\,\,{\rm fm}^{2},\qquad
c^\pi_S=11\pm 2\,\,{\rm GeV}^{-4}.
\ee
Notice that the result for $\langle r^2 \rangle^\pi_S$ is
also compatible with the lattice result of \cite{Aoki:2009qn}.

This is all the information we can extract from the scalar
formfactors. Currently, there are basically no results available for
$F^\pi_s(0)$ and for the energy behaviour of the kaon scalar formfactors or of
the strange contribution to the pion formfactor.

\subsection{$L^r_9$ and $L^r_{10}$}

We do not attempt to fit the remaining NLO LECs,  $L^r_9$ and $L^r_{10}$. 
Those LECs we fit are independent of $L^r_9$ and $L_{10}^r$, or alternatively,
none of the observables\footnote{with the exception of $K_{\ell4}$
 where a very small dependence is present for $s_\ell\ne0$.} 
we discuss depend on them.
One needs to include additional information to constrain their values.

$L^r_9$ appears alone at NLO in the electromagnetic radius of the pion vector
formfactor. The NNLO contribution dependent on the other $L_i^r$
is rather small  \cite{Bijnens:2002hp}.
It was therefore possible to fit that constant almost
independently from the other couplings \cite{Bijnens:2002hp}. 
Furthermore it never appears at NLO in any of the
observables used here as input thus it does not affect much our fits. We
always set  $L^r_9=(0.593\pm0.43)\times 10^{-2}$ for $\mu=0.77 \,\,{\rm GeV}$.

$L^r_{10}$ can be estimated using $\tau$ decays data on the $V-A$ spectral
function \cite{GonzalezAlonso:2008rf}. Its value was found to be
$L^r_{10}=(-4.06\pm0.39)\times10^{-3}$ at $\mu=0.77 \,\,{\rm GeV}$. 
However this constant never appears in the observables under
study not even at NNLO. Therefore it does not have any influence on our fits.  
For this reason we always set such constant to zero in our fits.

\section{Resonance estimates for the $C^{r}_i$}
\label{sect:Cires}

The many unknown coupling constants that appear in the $p^6$ Lagrangian,
the $C_i^r$,
represent the major problem for performing the fit with a $\mathcal{O}(p^6)$
 precision. A lot of effort went into trying to estimate them using
different models and treatments. The one we present here, also used in
\cite{Amoros:2000mc}, is the
resonance saturation model \cite{Ecker:1988te,Donoghue:1988ed}. 
It is based on the idea that the LECs encode the
information from physics above $\Lambda_{\rm ChPT}\approx
1$ GeV, and that they are dominated by the physics just above
this scale,
i.e. the physics of low-lying resonances. Therefore we need a Lagrangian
that describes these new particles and their interactions with the
pseudoscalar mesons of the octet. We include only vector, scalar and the
$\eta'$ fields. We use the same estimate described in
\cite{Amoros:2000mc}, thus we refer the reader to that paper for further
details, including the Lagrangians used at the resonance level.

The model is used then to estimate the $p^6$ contributions
depending on the $C^{r}_i$.
In \cite{Amoros:2000mc}, the heavier mesons were integrated out
producing $p^6$ Lagrangians for the pseudo-Goldstone boson. 
The heavy
resonance fields for the vector mesons produce
\ba
\label{LagInt}
\mathcal{L}_V & = & - \frac{i f_\chi g_V}{\sqrt{2} M^2_V} 
\langle \nabla_\lambda ([ u^\lambda, u^\nu ] ) [ u^\nu, \chi_- ] \rangle 
+ \frac{g_V \alpha_V}{\sqrt{2} M^2_V} 
\langle [ u_\lambda, f_-^{\nu \lambda} ] 
( \nabla^\mu [ u_\mu,u_\nu ] )\rangle \nonumber \\ 
& & - \frac{i g_V f_V}{2 M^2_V} \langle ( \nabla_\lambda f_+^{\lambda \nu} )
(\nabla^\mu [ u_\mu, u_\nu ] ) \rangle 
- \frac{i \alpha_V f_\chi}{M^2_V} \langle [ u_\nu, \chi_- ] 
[ u_\lambda,f_-^{\nu \lambda} ] \rangle \nonumber \\
& & - \frac{f_\chi f_V}{\sqrt{2} M^2_V} 
\langle ( \nabla_\lambda f_+^{\lambda \mu} ) 
[ u_\mu, \chi_- ] \rangle \; ,
\ea
and the scalar mesons
\ba
\label{LagInt2}
\mathcal{L}_S & = & \frac{c_d^2}{2 M^4_S} \langle 
\nabla_\nu (u_\mu u^\mu) \nabla^\nu (u_\lambda u^\lambda) \rangle 
+ \frac{c^2_m}{2 M^4_S} \langle ( \nabla_\nu \chi_+) 
( \nabla^\nu \chi_+ ) \rangle 
+ \frac{c_d c_m}{ M^4_S} \langle \nabla_\nu (u_\mu u^\mu) 
( \nabla^\nu \chi_+ ) \rangle \, . \nonumber\\
\ea
While for the $\eta'$ they obtained
\be
\label{etaprime}
\mathcal{L}_{\eta'} =  
- \frac{\tilde{d}_m^2}{2 M^4_{\eta'}} 
\partial_\mu \langle \chi_- \rangle \partial^\mu \langle \chi_- \rangle 
\ee
In (\ref{LagInt}) and (\ref{LagInt2}) $f^{\mu \nu}_{\pm}$ are defined as
\ba
f^{\mu \nu}_{\pm}  =  
u (v^{\mu \nu}-a^{\mu \nu})  u^\dag 
\pm u^\dag (v^{\mu \nu}+a^{\mu \nu}) u \nonumber \,.
\ea
In \cite{Amoros:2000mc} the above Lagrangians were not rewritten in the
standard form of the Lagrangian at $p^6$.
That work has since been done using more general resonance lagrangians
in   \cite{Cirigliano:2006hb,Kampf:2006yf}. We have checked that the
results using the Lagrangians (\ref{LagInt},\ref{LagInt2}) directly
agrees with the same inputs using the $C^r_i$ directly in terms of resonance
parameters as derived in  \cite{Cirigliano:2006hb,Kampf:2006yf}.
The $\eta^\prime$ contribution was rewritten in the $C_i^r$ in
\cite{eta3pi}.

The values we choose for the different couplings are the same
as in \cite{Amoros:2000mc}
\ba
\label{rescouplings}
&&f_V = 0.20,\quad  f_\chi = -0.025,\quad  g_V = 0.09, \nonumber \\
&&\alpha_V = -0.014,\quad c_m = 42 \mbox{ MeV},\quad
  c_d = 32 \mbox{ MeV},\nonumber\\
 &&\tilde{d}_m = 20 \, \, \mbox{MeV}.
\ea
and the masses are the experimental ones \cite{Nakamura:2010zzi}.
\be
m_V = m_\rho = 0.77 \mbox{ GeV},\qquad m_S = 0.98 \mbox{ GeV}, 
\qquad m_{\eta'} = 0.958 \mbox{ GeV},
\ee

In Table~\ref{tab:Ci} in the appendix we quote the $C^{r}_i$ as estimated
through the resonance model. We did not include more sophisticated resonance
models because this would have again increased strongly the number of free
parameters to be fitted. As discussed below we also have indications that
terms suppressed by $1/N_c$, $N_c$ the number of colours, might be important.
These cannot at present be estimated using this type of approach.
 
\section{Existing fits}
\label{sect:fit10}

In this section we describe a bit more in detail the earlier fits.
The main full fit done is fit 10 in  \cite{Amoros:2001cp}\footnote{
The E865 data were still
preliminary then, the main fit in \cite{Amoros:2001cp} was with older
$K_{\ell4}$ data.}. 
Earlier determination of the $L_i^r$ did not fully include NNLO effects
and we thus do not discuss them here.
The values for the $L_i^r$ obtained
in fit~10 are reproduced in Table~\ref{tab:fit10} in the column labelled
fit~10.
This is a full NNLO fit of the $L^{r}_i$
 and it
was done including the quantities and the $L_i^r$ whose value they
influence most:
\begin{table}
\begin{center}
\begin{tabular}{|c|cc|}
\hline
             & fit 10 &fit 10 iso   \\
 \hline
$10^3 L_1^r$ & $0.43$ &$0.39\pm0.12$ \\
$10^3 L_2^r$ & $0.73$&$0.73\pm0.12$   \\
$10^3 L_3^r$ &$ -2.35$&$-2.34\pm0.37$ \\
$10^3 L_4^r$ &$\equiv0$ &$\equiv0$ \\
$10^3 L_5^r$ &$0.97$ &$0.97\pm0.11$ \\
$10^3 L_6^r$ &$\equiv0$ &$\equiv 0$ \\
$10^3 L_7^r$ &$-0.31$ &$-0.30\pm0.15$ \\
$10^3 L_8^r$ &$0.60$ &$0.60\pm0.20$ \\
\hline
$\chi^2$ (dof) & - - & $0.26$ (1) \\
\hline
\end{tabular}
\caption{\label{tab:fit10} The results
  for fit10
  of \cite{Amoros:2000mc} and for a similar fit done without including isospin
  breaking corrections for the masses (fit10 iso) and also using the masses in
  (\ref{masses}) and decay constant $F_\pi$ as in (\ref{decayconst}). 
  The uncertainties are those calculated by MINUIT. The two fits reported are
  in agreement within uncertainties.}
\end{center}
\end{table}
\begin{table}
\begin{center}
\begin{tabular}{|c|ccc|}
\hline
 & $p^{2}$&$ p^{4}$& $p^{6}$\\
\hline
$m^{2}_{\pi}$ & $0.753$ & $0.006$ & $0.241$\\
$m^{2}_{K}$   & $0.702$ & $0.007$ & $0.291$\\
$m^{2}_{\eta}$& $0.747$ & $-0.047$& $0.300$\\
$F_\pi/F_0$       &$1$ & $0.136$ & $-0.075$\\
$F_K/F_0$         &$1$ & $0.307$ & $-0.003$\\
$F_K/F_\pi$   &$1$ & $0.171$& $0.049$ \\
\hline
\end{tabular}
\caption{\label{tab:fit10b}
 The convergence of the expansion for the meson masses and the decay
  constants for fit 10 iso.  A similar behaviour holds for fit10. The masses
  quoted are normalized to the physical masses, while the decay constants to
  $F_0$ ($F_0=0.0869$ GeV).}
\end{center}
\end{table}
\begin{enumerate}
\item masses and pion decay constant with the old values as in
  (\ref{massesold}) and (\ref{decayconstold}) 
\item the $K_{\ell 4}$ formfactor parameters: 
$f_{s}$, $f'_{s}$, $g_{s}$, $g'_{s}$. They constrained mostly 
$L^{r}_{1},L^{r}_{2},L^{r}_{3}$.
\item $F_{K}/F_{\pi}=1.22\pm 0.01$, sensitive to $L^{r}_{5}$.
\item $m_{s}/\hat{m}=24$ constrains
$L^{r}_{7},L^{r}_{8}$ via
the masses in (\ref{massratio}).
\item $L^{r}_{4}\equiv L^{r}_{6} \equiv 0$  since they are $1/N_{c}$
suppressed couplings.
\end{enumerate}

The $C^{r}_{i}$ contributions were estimated using resonance saturation as
described in Section~\ref{sect:Cires}. They also included there the
axial-vector resonances, although their contribution was rather small. 
The scale of saturation was set to $\mu\equiv0.77\,{\rm GeV}$, 
but $\mu=0.5,\,1\,{\rm  GeV}$ were within errors.
In fit 10 isospin breaking corrections in the masses and
decay constants  were also included, though 
the authors of \cite{Amoros:2001cp} noticed that the
neglect of isospin violation was a good approximation. Indeed the fits
performed including or not these effects are in agreement within errors as can
be seen from Table~\ref{tab:fit10} comparing the columns fit 10 and fit 10 iso.

Fit 10 has been so far a quite successful fit. Not only because it already
included many
quantities at order $p^6$, but also because the resulting $L^{r}_i$ nicely
confirmed the estimates from
resonance models. These are lead by the large $N_c$ expansion which
predicts e.g. $2L^{r}_1\approx L^{r}_2$
 and $L^{r}_4\approx L^{r}_6 \approx 0$. While the second
relation was imposed, the first one was found to be well
satisfied. This added credibility to the fit itself even
though it relied on the resonance estimate for the tree-level $p^6$ 
contributions.

However the convergence of the perturbative expansion for this fit is not as
expected.
The different orders for the masses and decay constants
are reported in Table~\ref{tab:fit10b} for fit 10 iso\footnote{The numbers for
fit 10 itself can be found in Table~2 of \cite{Bijnens:2003xg}.}.
The $\mathcal{O}(p^4)$ order of the masses turns out to be tiny, far less
than the expected $30\%$. On the other hand the NNLO contribution is
definitely too large. The sources of this bad convergence are
basically two. First the constraint $L^{r}_{4}\equiv L^{r}_{6}\equiv 0$ that
clearly sends to zero many contributions coming from the NLO tree-level
diagrams. Secondly most of
the $C^{r}_{i}$ appearing in the masses expressions are estimated to be
zero as well. Therefore they cannot help in
canceling  large two-loop contributions.
On the other hand the convergence for the decay constants is quite satisfying. 

After fit 10 was performed many other observables have been calculated at
$\mathcal{O}(p^6)$ in $SU(3)$ ChPT such as the $\pi\pi$ and $\pi K$ scattering
threshold parameters. Of course it is very important to compare the pure ChPT
predictions obtained
using fit~10 with the values of Tables~\ref{tab:pipi} and \ref{tab:piK}. These
comparisons have been done and can be found in Table~1 of \cite{BDT} and in
Table~4 of \cite{BDT2}. Fit~10 is mostly in agreement within errors, although
there are small discrepancies in some of the threshold parameters.
Some comparison with scalar formfactors was done in \cite{Bijnens:2003xg}.
The last three papers used the same inputs as fit~10 and tried to
vary $L_4^r$ and $L_6^r$ to see if some preferred regions could be
found. Here we redo the fit from the beginning with all inputs.

\section{New Fits}
\label{sect:newfits}

The aim of this section is to show how the new measurements and
observables included in our global fits, change the results
compared to fit 10. We have rewritten as mentioned above all programs
into C++ and are
using the isospin symmetric versions of the calculations. 
We therefore first redid the fit using the same 
inputs as fit 10. The outputs are given in Table~\ref{tab:fit10} in the column
labelled fit 10 iso. This also shows that the minor changes in masses
and $F_\pi$ as well the isospin breaking corrections do not affect the 
fit values appreciably. We will now add the effects of the changed
experimental inputs and of the additional inputs to see how
they change the fitted values of the $L_i^r$.

In Table~\ref{tab:fitall} we present several fits. 
These have all been performed using
the resonance estimate of the $C^r_i$ of Section~\ref{sect:Cires} and
Table~\ref{tab:Ci} in the appendix, setting the scale of saturation
$\mu=0.77$ GeV. Furthermore, we used the new values of the masses
and decay constant of (\ref{masses}) and (\ref{decayconst}). 
We remind the reader that the use of these new parameter-values affects
the output only within the uncertainties.
\begin{table}
\begin{center}
\begin{tabular}{|l|c c c c |c| c c |}
\hline
             &fit 10 iso    & NA48/2   & $F_K/F_\pi$ & All $\star$ &  All  & $C^r_i\equiv0$& All $p^4$\\ \hline
$10^3 L_1^r$ &$0.39\pm0.12$ & $0.88$    & $0.87$     &$0.89$  &$0.88\pm0.09$ & $0.65$ &$1.12$\\
$10^3 L_2^r$ &$0.73\pm0.12$ & $0.79$    & $0.80$     &$0.63$  &$0.61\pm0.20$ & $0.11$ &$1.23$\\
$10^3 L_3^r$ &$-2.34\pm0.37$& $-3.11$   &$-3.09$     &$-3.06$ &$-3.04\pm0.43$& $-1.47$&$-3.98$\\
$10^3 L_4^r$ &$\equiv0$     & $\equiv0$ & $\equiv0$  & $0.60$ &$0.75\pm0.75$ & $0.80$ &$1.50$\\
$10^3 L_5^r$ &$0.97\pm0.11$ & $0.91$     &  $0.73$   &$0.58$  &$0.58\pm0.13$ & $0.68$ &$1.21$\\
$10^3 L_6^r$ &$\equiv 0$    & $\equiv 0$& $\equiv 0$ & $0.08$ &$0.29\pm0.85$ & $0.29$ &$1.17$   \\
$10^3 L_7^r$ &$-0.30\pm0.15$& $-0.30$    &$-0.26$    &$-0.22$ &$-0.11\pm0.15$& $-0.14$&$-0.36$  \\
$10^3 L_8^r$ &$0.60\pm0.20$ & $0.59$     & $0.49$    &$0.40$  &$0.18\pm0.18$ & $0.19$ &$0.62$  \\ \hline
$\chi^2$  & $0.26$  &$0.01$  &$0.01$   &$1.20$  &$1.28$  &$1.67$ &$2.60$ \\
dof  & 1 & 1 & 1 & 4 & 4 & 4 & 4\\
\hline
\end{tabular}
\caption{\label{tab:fitall}Several global fits compared to fit10 iso. For all
  the fit $\mu=0.77$ GeV. The errors quoted are the ones as calculated by
  MINUIT. The numbers in the row labelled dof are the degrees of freedom for the fit. See the
  description in the text for further details on how the fits have been performed.
  The column labelled All is our main new result.}
\end{center}
\end{table}                                            
Hereafter we summarize the steps in which we have included the new information.
\begin{description}
\item[{\rm NA48/2}] The input observables and their values are the same as for
  fit 10 iso, but we use the
  new measurements in (\ref{kl4newmeas}) 
  for the $K_{\ell 4}$ decay from the NA48/2
  collaboration \cite{Batley:2010zza}. 
  The new measurements lead immediately to a striking feature:
  the large $N_c$ relation $2 L^{r}_1\approx L^{r}_2 $ does not hold any
  longer. It even turns out that $L^{r}_2 \lesssim L^{r}_1$. Notice that,
  as explained in
  Section~\ref{sect:kl4}, the slope $f'_s$ comes from a second
  order fit of the $f_s$ formfactor and therefore it differs from the one used
  in fit 10. In Section~\ref{sect:linfit} we will present also results for the
  linear fit of the $f_s$ formfactor.
\item[$F_K/F_\pi$] Same as fit NA48/2 but with the new value in (\ref{ratio})
  for $F_K/F_\pi$. $L^{r}_5$ is mainly affected and becomes smaller than in
  fit 10. As a consequence also the
  convergence of the decay constants expansion is
  worsened, e.g. $F_\pi/F_0|_{p^4}\approx 0.134$ while
  $F_\pi/F_0|_{p^6}\approx-0.126$.
\item[{\rm All$\star$}] In this fit we include a few more observables, i.e. the
  $\pi\pi$ scattering parameters $a^0_0$ and $a^0_2$, the $\pi K$
  scattering parameters $a^{\frac{1}{2}}_0$ and $a^{\frac{3}{2}}_0$ and the
  pion scalar radius $\langle r \rangle^2_S$. We also release the constraint
  $L^{r}_4=L^{r}_6=0$ because now we have many more observables included. 
  Unfortunately
  since none of the new observables involve information on the masses (and
  therefore on those two couplings), we still cannot achieve the
  precise values of $L^{r}_4$ and $L^{r}_6$. They are
  highly correlated and MINUIT gives a very large uncertainty for
  those. Furthermore since now the $\mathcal{O}(p^4)$ contributions to the
  masses due to $L^{r}_4$ and $L^{r}_6$ are not zero, $L^{r}_5$ and $L^{r}_8$
  diminish.
\item[{\rm All}] This fit is very similar to fit All$\star$. 
  Here we adopt the new value for the quark mass ratio
  $m_s/\hat{m}=27.8$. However we find that using values for the quark mass ratio
   between $27$ and $29$ does not
  change the results considerably. 
  The constants $L^{r}_7$, that appears in the $\eta$ mass, and $L^{r}_8$ are
  strongly affected by this change. This is also relatively true for $L^{r}_4$
  and $L^{r}_6$.
  We also tried to perform the same fit but setting $L^{r}_4\equiv L^{r}_6\equiv
  0$. The resulting fit is very similar to fit NA48/2 but it has a huge
  $\chi^2$ ($\chi^2=45$). 
\item[$C^r_i\equiv0$] In the last column of the table we quote the fit
  obtained including the same input as for fit All, but setting all the
  $C^r_i\equiv 0$.
 This fit has been done to show how the different $C^r_i$ can affect the
 $L^r_i$ fit. Notice that the
  constants $L^r_1$, $L^r_2$ and $L^r_3$ change a lot, while the others
  stay in the same area as in fit All. This is not surprising: the last few constants
  are indeed primarly fitted from quantities where many contributing $C^r_i$
  are large-$N_c$ suppressed and those which are not are set to zero
  also in the simple resonance estimate used.
\item[{\rm All $p^4$}] Same fit as All but all expressions are now at NLO. Use this fit
for one-loop ChPT results. Note that this produces very high values for $L_4^r$ and
$L_6^r$. The underlying reason is that the lower value of $F_K/F_\pi$ requires a smaller
$L_5^r$ than before and the pion scalar radius then requires at this order a larger
$L_4^r$. This effect is also visible in fit All but is reduced when including the NNLO corrections.
\end{description}

Fit All is what we consider as the present best fit for NNLO ChPT calculations,
it thus superseeds fit 10 of \cite{Amoros:2001cp}.

Let us discuss how the ChPT expansion is affected by the new values for the
$L_i^r$.
The various terms of the mass expansions read for fit All
\ba
\label{massconvfitAll}
m^2_\pi|_{p^2}=1.035 \quad&
m^2_\pi|_{p^4}=-0.084 & \quad
m^2_\pi|_{p^6}=+0.049\,, \nonumber\\
m^2_K|_{p^2}=1.106 \quad&
m^2_K|_{p^4}=-0.181 &\quad
m^2_K|_{p^6}=+0.075\,,\\
m^2_\eta|_{p^2}=1.186 \quad&
m^2_\eta|_{p^4}=-0.224 &\quad
m^2_\eta|_{p^6}=+0.038\,,\nonumber
\ea
while those for the decay constants are
\ba\label{decayconvfitAll}
&&\left.\frac{F_\pi}{F_0}\right|_{p^4}=0.311 
\quad\left.\frac{F_\pi}{F_0}\right|_{p^6}=0.108\nonumber\\
&&\left.\frac{F_K}{F_0}\right|_{p^4}=0.441 
\quad\left.\frac{F_K}{F_0}\right|_{p^6}=0.216\,,\\
&&\left.\frac{F_K}{F_\pi}\right|_{p^4}=0.129
\quad\left.\frac{F_K}{F_\pi}\right|_{p^6}=0.068\,. \nonumber
\ea
In (\ref{massconvfitAll}) and (\ref{decayconvfitAll}) we used the same
normalizations as in Table~\ref{tab:fit10b}, although now $F_0=0.065$ GeV,
this is due to the larger value of $L_4^r$ which comes however with
a large error. 
 Notice that the convergence of the
mass expansions in (\ref{massconvfitAll}) is improved  compared to the one of
fit 10 in Table~\ref{tab:fit10b}. However (\ref{massconvfitAll}) looks
strange: the LO masses are larger than the physical ones and there are significant
cancellations between NLO and NNLO.
Furthermore, even if the convergence is
improved, it is still quite different from the one expected. 
E.g. the $m_\pi^2|_{p^4}$
contribution is much smaller than the expected $30\%$ and it is of the same size
as the $p^6$ order. The convergence for the
decay constants is a bit worsened compared to the one of fit 10, due
to the low value of $L^{r}_5$, but it is still acceptable. Notice also that
when the ratio $F_K/F_\pi$ is calculated with (\ref{ratio2}) the resulting
value is $1.168$, which is $3\%$ smaller than the expected $1.197$ . This can
be due to higher order corrections that are included in the ratio of
(\ref{ratio2}), but not in (\ref{ratio1}).

We performed more fits than those quoted in Table~\ref{tab:fitall}.
We included more $\pi\pi$ scattering parameters and $\pi K$
scattering parameters. We found that these fits are compatible with fit All
of Table~\ref{tab:fitall} within uncertainties. The same is true when we add
the quantity $c^\pi_s$.

For completeness we quote the value for $f_+(0)$, the formfactor of
$K_{\ell3}$ decay at zero momentum transfer.  This quantity was calculated at
NNLO in \cite{Bijnens:2003uy}. We check how much the new
$L^r_i$ of fit
All would affect its value. Notice that at zero momentum the dependence on
$L^r_9$ drops out. The other $L^r_i$ appear only at NNLO. 
We discuss both the
case of the charged kaon, the $K^+_{\ell3}$ decay, and the
neutral one $K^0_{\ell3}$. 
The results we obtain using the masses of the particles involved in the decay
are given in Table \ref{tab:kl3}. These numbers can be seen as an update of
those in Table 3 of \cite{Bijnens:2003uy}.
\begin{table}
\begin{center}
\begin{tabular}{|r|cc|}
\hline
                       & $K^0_{\ell3}$ & $K^+_{\ell3}$ \\
\hline
$p^4$                  & $-$0.02276 & $-$0.02287\\
$p^6$ loops only       &    0.1141  &    0.01115\\
$p^6$-$L_i^r$ fit 10 iso &    0.00342 &    0.00330\\
$p^6$-$L_i^r$ fit All    &    0.00232 &    0.00247\\
\hline
\end{tabular}
\end{center}
\caption{\label{tab:kl3} The results for $f_+(0)$ in the two $K_{\ell3}$ decays.
This is an update of Table 3 in \cite{Bijnens:2003uy}}
\end{table}
The numbers for the pure loops are changed w.r.t. \cite{Bijnens:2003uy}
mainly because of the change in $F_\pi$. Fit 10 iso is essentially the same as 
fit 10 used in \cite{Bijnens:2003uy} but has a small (1\%) difference.
The $L_i^r$-dependent contribution changes but since it stays small the total
result is very similar to \cite{Bijnens:2003uy}.
The total value for $f_+(0)$ for $K^+_{\ell3}$ is instead 
\be\label{fpp}
\left. f_+(0)\right|_{{\rm C_i=0}(L^r_i={\rm fit 10})}=0.9916,\qquad
\left. f_+(0)\right|_{{\rm C_i=0}(L^r_i= {\rm fit All})}=0.9908
\ee
and for $f_+(0)$ for $K^0_{\ell3}$ is instead 
\be\label{fp0}
\left. f_+(0)\right|_{{\rm C_i=0}(L^r_i={\rm fit 10})}=0.9921,\qquad
\left. f_+(0)\right|_{{\rm C_i=0}(L^r_i= {\rm fit All})}=0.9910.
\ee
By comparing the two results in (\ref{fpp}) and in (\ref{fp0}) one
realizes that the differences due to the $L^r_i$ have a very small effect on
$f_+(0)$.

The main uncertainty still remains the value of the contribution of the
$C_i$. The estimate for the relevant constants used in this paper
leads to $\left.f_+(0)\right|_{C_i}\approx -0.045$ but the fitting inputs
used here do not strongly constrain the relevant combination.

\subsection{Linear fit for $K_{\ell 4}$ decays}
\label{sect:linfit}

One of the most striking features of the results presented in
Table~\ref{tab:fitall} is that as soon as the new results from the quadratic
fit of the NA48/2 collaboration \cite{Batley:2010zza} are included, 
the constants $L^{r}_1$ and
$L^{r}_2$ take unexpected values. Indeed, as was noted above,  they do not
respect any longer the large-$N_c$ relation $2L^r_1\approx L^r_2$, but already in
fit NA48/2 they are $L^{r}_1\approx L^{r}_2$ while when also the $\pi\pi$ 
and the $\pi K$ scattering lengths are included (fit All) we even obtain 
$L^{r}_1>L^{r}_2$.
On the other hand when we calculate the curvature $f''_s$ using the $L^{r}_i$ as
obtained in fit All we obtain $f''_s=-0.124$ to be contrasted with the
experimental value $f''_s=-0.437$. Furthermore, whenever we include as input
also $f''_s$ we again obtain fits compatible to the ones in
Table~\ref{tab:fitall}, but with much larger $\chi^2$ (e.g. $\chi^2\sim35$ for
fit All) and the largest contribution comes exactly from $f''_s$. These
results confirm what was already stated at the end of Section~\ref{sect:kl4},
i.e. the state-of-art ChPT does not reproduce such a large negative
bend. Since $f'_s$ and $f''_s$ are highly correlated, the linear and the quadratic
fit of the $F_s$ formfactor present rather different slopes.

For such reasons we perform fits using the slope of the linear fit
$f'_s/f_s=0.073$ \cite{Batley:2010zza} as well. The resulting fit, analogous to
fit All, is reported in Table~\ref{tab:fitalllin}.
\begin{table}
\begin{center}
\begin{tabular}{|c|c|}
\hline
             & All linear\\
\hline
$10^3 L^r_1 $&$ 0.58\pm0.10$\\
$10^3 L^r_2 $&$ 0.80\pm0.12$\\
$10^3 L^r_3 $&$-3.33\pm1.42$\\
$10^3 L^r_4 $&$0.93\pm0.31$\\
$10^3 L^r_5 $&$ 0.71\pm0.24$\\
$10^3 L^r_6 $&$0.86\pm0.87$\\
$10^3 L^r_7 $&$-0.04\pm0.40$\\
$10^3 L^r_8 $&$ 0.02\pm0.79$\\
\hline
$\chi^2$(dof)&$1.16 (4)$\\
\hline
\end{tabular}
\end{center}
\caption{\label{tab:fitalllin} The analogous of fit All, but the linear fit
  of the $K_{\ell 4}$ formfactors have been used. The values of the $L^r_i$
  are at $\mu=0.77$ GeV.}
\end{table}
By inspection one can see that the large $N_c$
relation $2L^r_1=L^r_2$ still does not precisely hold, but at least $1.4
L^r_1\approx L^r_2$. On the other hand $L^r_4$ and $L^r_6$ are again not
suppressed, while $L^r_7$ and $L^r_8$ are unexpectedly small. Also
all the constants have large uncertainties.

The convergence of the chiral expansions is worse than the
one for fit All. The various terms of the mass expansions read 
\ba\label{massconvfitAlllin}
m^2_\pi|_{p^2}=0.655 \quad&
m^2_\pi|_{p^4}=0.370 & \quad
m^2_\pi|_{p^6}=-0.025\,, \nonumber\\
m^2_K|_{p^2}=0.699 \quad&
m^2_K|_{p^4}=0.181 &\quad
m^2_K|_{p^6}=0.120\,,\\
m^2_\eta|_{p^2}=0.751 \quad&
m^2_\eta|_{p^4}=0.151 &\quad
m^2_\eta|_{p^6}=0.098\,,\nonumber
\ea
while those for the decay constants are
\ba\label{decayconvfitAlllin}
&&\left.\frac{F_\pi}{F_0}\right|_{p^4}=0.355 
\quad\left.\frac{F_\pi}{F_0}\right|_{p^6}=0.157\nonumber\\
&&\left.\frac{F_K}{F_0}\right|_{p^4}=0.498 
\quad\left.\frac{F_K}{F_0}\right|_{p^6}=0.262\,,\\
&&\left.\frac{F_K}{F_\pi}\right|_{p^4}=0.143
\quad\left.\frac{F_K}{F_\pi}\right|_{p^6}=0.054\,. \nonumber
\ea

In the light of these results it is rather difficult to draw a
conclusion. The very different predictions for $L^r_1$ and $L^r_2$
obtained in fit All and this fit confirm that the picture of ChPT
for $K_{\ell 4}$ decays is still incomplete. As mentioned earlier, we expect a
dispersive analysis to produce a larger curvature.

\subsection{Some small variations on fit All}
\label{sect:confres}

In the resonance estimate described in Section~\ref{sect:Cires} there is at
least an assumption not entirely justified. We assume the scale at
which the saturation happens to be $0.77$ GeV, i.e. the mass of the lowest
lying resonance. Nothing prevents us to choose a larger or smaller
scale, although this is still expected to be in the same range of energy. 
To check whether this
assumption is safe we try to fit from data the saturation scale parameter as
well. The results are rather reassuring. 
Fit All of Table~\ref{tab:fitall} is completely 
unaffected by this procedure. The fitted
saturation scale is $0.77\pm0.45$ GeV.

The fit in Table~\ref{tab:fitalllin} shows a little difference. The fitted
saturation scale is now $0.71\pm0.31$ GeV. However the $L^r_i$ do not change that
much and the look of the fit is pretty much the same as before.

We also attempt to find better estimates of the $C^r_i$ constants
releasing the values of the couplings $g_V$, $c_d$ and $c_m$. Again
we try to fit them using both the input of fit All and of the fit in
Table~\ref{tab:fitalllin} (linear fit of NA48/2 instead of quadratic). 
In the first case we find in fact $g_V$ and $c_m$ close to the ones in
(\ref{rescouplings}). They read $g_V=0.097\pm0.123$ and $c_m=0.045\pm0.049$~GeV. For $c_d$ we find
instead a value larger than expected, i.e. $c_d=0.093\pm0.100~$GeV. Anyway they are all 
affected by large uncertainties. The $L^r_i$ fit is somewhat compatible with
the one of Table~\ref{tab:fitall}, fit All, within uncertainities because of the large ranges
allowed for $g_V,c_m,c_d$. The fits are in a very broad minimum here
with only one degree of freedom.

If we apply the same procedure but with the same
$K_{\ell4}$ input as for Table~\ref{tab:fitalllin} (NA48/2 linear fit)
we arrive to similar conclusions: the values
of $g_V$ and $c_d$ are similar to the ones in (\ref{rescouplings}), while $c_d$
is larger. Again all the resonance couplings present large uncertainties.
The $L_i^r$ are here rather well compatible with those in Table~\ref{tab:fitalllin}.

We also try to multiply the $C^r_i$ by an overall constant $\alpha$ and
include it as a fitting parameter. It is encouraging to see that the result is
$\alpha\approx 1.03$, namely the best fit is reached with basically the same
values of the $C^r_i$ from resonance estimate. Obviously the fit obtained is
very similar to fit All.
When we apply the same procedure to the fit in Table~\ref{tab:fitalllin} the
constant $\alpha$ takes the value $0.90$. This affects the fit of
Table~\ref{tab:fitalllin}, but still within uncertanties.

{}From this we conclude that fit All is stable against small changes in the resonance
estimate of the $C_i^r$.

\subsection{Adding input: $\bar{\ell_i}$ constants in two-flavour ChPT}
\label{sect:li}

The authors of \cite{Gasser:2007sg,Gasser:2009hr} study three flavour ChPT in
the limit where the $m_s$ is assumed to be much larger than $\hat{m}$
and the external momenta.  In this case
they can integrate out the strange quarks and $SU(3)\times SU(3)$ ChPT reduces
to $SU(2)\times SU(2)$ ChPT. Matching the results from the two frameworks 
they calculate explicitely the dependence of the two-flavour LECs (the
scale-independent $\bar{\ell_i}$ and the $c^r_i$) on the
strange quark mass and on the three-flavour LECs. These relations have been worked
out using two different methods at order $p^6$ in
\cite{Gasser:2007sg,Gasser:2009hr}.
 
There exist different evaluations of the $\bar{\ell_i}$. 
$\bar{\ell_3}$ has been estimated
rather well using lattice results \cite{Colangelo:2010et}. $\bar{\ell_1}$,
$\bar{\ell_2}$, $\bar{\ell_4}$ and $\bar{\ell_6}$ have been obtained by
matching two-flavour ChPT with dispersive results \cite{CGL}, but
(contradictory) lattice results exist for those too \cite{Colangelo:2010et}.
We have increased the error on $\bar\ell_4$ because of this.
\begin{table}
\begin{center}
\begin{tabular}{|l|cccc|}
\hline
&fit 10 iso
&All
&All linear
&\cite{CGL,Colangelo:2010et} \\
\hline
$\bar{\ell_{1}}$&$-0.6 (0.5)$ &$-0.1 (1.1)$&$-1.9$& $-0.4\pm0.6$\\
$\bar{\ell_{2}}$&$5.7 (4.9)$&$5.3 (4.6)$&$5.7$&$4.3\pm0.1$\\
$\bar{\ell_{3}}$&$1.3 (2.9)$&$4.2 (4.9)$&$4.1$&$3.3\pm0.7$\\
$\bar{\ell_{4}}$&$4.0 (4.1)$&$4.8 (4.8)$&$4.5$&$4.4\pm0.4$\\
\hline
\end{tabular}
\end{center}
\caption{\label{tab:li} The values of the scale-independent $SU(2)$ LECs
 $\bar{\ell_i}$. In the
  first three columns we show the values as predicted by fit 10, fit All and fit All linear using the NNLO matching conditions of  \cite{Gasser:2007sg}. The numbers between parenthesis are the NLO results.  In
  the last column we quote the known values from
  \cite{CGL,Colangelo:2010et}. Notice that the uncertainty over $\bar{\ell_4}$ is
  double the one quoted in \cite{CGL} due to the still unclear situation for
  the lattice results \cite{Colangelo:2010et}.}
\end{table}

In Table~\ref{tab:li} we summarize all the values of these constants and the
results obtained by plugging the 
$L^r_i$ and $C^r_i$ of fit 10 iso, fit All and fit All linear in both the NNLO and NLO relations
of \cite{Gasser:2007sg,Gasser:2009hr}. By comparison of the first three
columns with the last one of
Table~\ref{tab:li} it is easy to see that none of the fits correctly
reproduces all the $\bar{\ell_i}$ values. A similar conclusion holds also for
the fit in the next-to-last column of Table~\ref{tab:fitall}, where all the $C^r_i$
are zero. The disagreement with the $\bar{\ell_i}$ in this case is actually
even stronger. These fits
encounter particular trouble in
fitting $\bar{\ell_2}$. 

We tried to fit the $\bar{\ell_i}$ whose values appear on the last column of
Table~\ref{tab:li} in addition to the inputs used for fit All.
Not surprisingly it is not possible to accommodate all those inputs
at the same time. The resulting $\chi^2$ is
approximately $22$ and its largest contributions come exactly from the
$\bar{\ell_i}$. Excluding from the fit $\bar{\ell_2}$
but including the others improves
the situation. The resulting $L^{r}_i$ values are very close to the ones of fit
All, the most important deviation being $10^{3}L^r_3=-3.18$. The $\chi^2$
takes the value $3.15$ with $7$ degrees of freedom. Also in this case
the value for $\bar{\ell_2}$ is still far from the expected one.

This is not surprising. In \cite{Gasser:2007sg} it was found that the constant
$\bar{\ell_2}$ depends on the couplings $L^r_2$, $L^r_3$ and on the
combination $2C^r_{13}-C^r_{11}$. The authors of  \cite{Gasser:2007sg}
observed there that to find agreement with the determined value of
$\bar{\ell_2}$ the combination of $C^r_i$ must not be zero. Unfortunately
those are two large-$N_c$ suppressed couplings and therefore they are set to
zero in our resonance estimate (see Table~\ref{tab:Ci}). We also try to fit
those two $C^r_i$ using also $\bar{\ell_2}$ as input observable, but this has
been unsuccessful as well. In this way we manage to accommodate the value for
$\bar{\ell_2}$, but then $\bar{\ell_1}$ is off, since it contains also a
different combination of $C^r_{11}$, $C^r_{13}$ and of $C^r_6$. This
last coupling is also $N_c$-suppressed and thus estimated to be zero. In the
end there is no way out: when we try to fit $C^r_6$ too, there are other
quantities taking very different values. The $C^r_i$ are too
correlated to be able to fit only a few of them.

As far as regards the fit in Table~\ref{tab:fitalllin} the results for the
$\bar{\ell_i}$ are even less clear. Its predictions are reported in the
third column of Table~\ref{tab:li}. It is straightforward to see that now
even the predicted $\bar{\ell_1}$ is off. Of course when we try to fit all the
$\bar{\ell_i}$ the $\chi^2$ is very large ($\chi^2\approx 37.7$). Contrary to
what happened for fit All, the situation does not improve
that much when we exclude $\bar{\ell_2}$. For this fit seems to be very hard
to reach the correct value for $\bar{\ell_1}$ too. The resulting $\chi^2$ in
this second case is $5.82$ with $\bar{\ell_1}=-1.4$.
 
Finally an extra cautionary remark. Requiring that the $SU(3)$ ChPT constants
predict the values for the $SU(2)$ ones might not be a very safe
assumption. What it assumes is that both $SU(2)$ and $SU(3)$ ChPT work well
for the same quantities. For the $\pi\pi$ scattering
quantities, which are very much determined by loop parts, relatively small differences
can become amplified in the resulting values of the LECs.

\subsection{A chiral quark model estimate for the $C^r_i$}
\label{sect:chinaCi}

We discussed above a simple resonance saturation estimate for the NNLO LECs $C_i^r$.
There are other attempts at predicting these values as well from
chiral quark models. As a representative of this we choose
 \cite{Jiang:2009uf}.
It also is a large $N_c$ approximation but with a somewhat different pattern
than our resonance saturation.
Their method is based on a study of the relation between  the chiral
Lagrangian up to order $p^6$ and QCD, they find as expected that the LECs can
be given in terms of some Green functions of QCD. In the evaluation 
of these Green functions, several
assumptions and approximations are made such that it is not a full derivation
but something like a chiral quark model.
Their results are presented in
Table~IV of \cite{Jiang:2009uf}.
\begin{table}
\begin{center}
\begin{tabular}{|c|cc|}
\hline
        &$C^r_i$\cite{Jiang:2009uf} 
&$\alpha \times C^r_i$\cite{Jiang:2009uf} \\
\hline
$10^3 L_1^r$ &  $0.66\pm0.11$    &$0.66\pm0.10$ \\
$10^3 L_2^r$ &  $0.59\pm0.13$    &$0.24\pm0.32$   \\
$10^3 L_3^r$ &  $-2.74\pm0.48$   &$-1.80\pm0.75$ \\
$10^3 L_4^r$ &  $0.75\pm0.16$    &$0.77\pm0.84$ \\
$10^3 L_5^r$ &  $1.64\pm0.83$    &$0.83\pm0.39$ \\
$10^3 L_6^r$ &  $0.64\pm0.41$    &$0.32\pm0.99$ \\
$10^3 L_7^r$ &  $-0.25\pm0.30$   &$-0.15\pm0.14$ \\
$10^3 L_8^r$ &  $0.76\pm0.75$    &$0.27\pm0.23$ \\
$\alpha$     &   --              &$0.27\pm0.47$ \\
\hline
$\chi^2$ (dof)& $3.71$ (4)       & $1.35$ (3) \\
\hline
\end{tabular}
\end{center}
\caption{\label{tab:chinaCi}The results as obtained using the $C^{r}_i$ estimates
  of \cite{Jiang:2009uf}. Both the fits include the same observables as fit
  All of Table~\ref{tab:fitall}. In the second column the coefficient $\alpha$
  multiplied by the $C^{r}_i$ has been included as fitting parameter. The $L^r_i$ are given at $\mu=0.77$
GeV.}
\end{table}
\begin{table}
\begin{center}
\begin{tabular}{|c|ccc|}
\hline
 & $p^{2}$&$ p^{4}$& $p^{6}$\\
\hline
$m^{2}_{\pi}$ & $0.988$ & $-0.066$ & $0.078$\\
$m^{2}_{K}$   & $1.056$ & $-0.177$ & $0.121$\\
$m^{2}_{\eta}$& $1.131$ & $-0.225$& $0.094$\\
$F_\pi/F_0$       &$1$ & $0.318$ & $0.108$\\
$F_K/F_0$         &$1$ & $0.475$ & $0.198$\\
$F_K/F_\pi$   &$1$ & $0.156$& $-0.050$ \\
\hline
\end{tabular}
\end{center}
\caption{\label{tab:chinaCib} The results as obtained using the $C^{r}_i$ estimates
  of \cite{Jiang:2009uf} showing the convergence for the fit where $\alpha$ is
left free. The normalizations are the same as explained in
Table~\ref{tab:fit10b}. Now $F_0=0.065$ GeV.}
\end{table}

We also use their estimate to perform the fits.
The results can be found in
Table~\ref{tab:chinaCi}. There are results for two
different fits. They have been obtained including all the observables as for
fit All of Table~\ref{tab:fitall}. In the first column we use the
$C^r_i$ as quoted in Table~IV of \cite{Jiang:2009uf}, whereas in the second
column we multiply them by an overall constant $\alpha$ that is also fitted. This
second fit was done because we observed that the values for the $C^{r}_i$ of
\cite{Jiang:2009uf} are somewhat larger than the ones of the resonance
estimates of Table~\ref{tab:Ci}. The fit confirms this observation and finds
as best value for $\alpha=0.27$, i.e. $C^{r}_i$ considerably smaller than the
ones in \cite{Jiang:2009uf}. The value of the $\chi^2$ for the two fits of
Table~\ref{tab:chinaCi} is somewhat worse than for fit All. Indeed it seems
that it is now very difficult to fit the slope $g'_p$ of the $G$ formfactor
for $K_{\ell 4}$ decay. From the second fit of Table~\ref{tab:chinaCi} one can
notice that when the $C^{r}_i$ are allowed to take
smaller values the $L^r_2$ constant compensates for that. This allows the fit
to reach a better value for the $g'_p$.
 
As can be seen in
Table~\ref{tab:chinaCib}, even the convergence for the masses and decay constants
is worse than the one for fit All reported in (\ref{massconvfitAll}) and
(\ref{decayconvfitAll}) respectively. Notice that we have not quoted the
convergence for the fit obtained without multiplying the $C^r_i$ by the
coefficient $\alpha$. In fact this is found to be even worse than the one of
Table~\ref{tab:chinaCi}, the $p^4$ terms being constantly larger than, although
comparable in size to, the $p^6$ ones. 
\begin{table}
\begin{center}
\begin{tabular}{|c|cc|}
\hline
        &$C^{r}_i$ \cite{Jiang:2009uf} 
&$\alpha \times C^{r}_i$ \cite{Jiang:2009uf} \\
\hline
$10^3 L_1^r$ &  $0.38\pm0.10$    &$0.35\pm0.11$ \\
$10^3 L_2^r$ &  $0.88\pm0.12$    &$0.43\pm0.30$   \\
$10^3 L_3^r$ &  $-3.20\pm0.47$   &$-2.04\pm0.79$ \\
$10^3 L_4^r$ &  $0.42\pm0.17$    &$0.91\pm0.35$ \\
$10^3 L_5^r$ &  $1.62\pm0.77$    &$1.03\pm0.57$ \\
$10^3 L_6^r$ &  $0.43\pm0.21$    &$0.87\pm0.77$ \\
$10^3 L_7^r$ &  $-0.32\pm0.45$   &$-0.11\pm0.34$ \\
$10^3 L_8^r$ &  $0.92\pm1.07$    &$0.17\pm0.71$ \\
$\alpha$     &   --             &$0.22\pm0.47$ \\
\hline
$\chi^2$ (dof)& $4.13$ (4)    & $1.20$ (3) \\
\hline
\end{tabular}
\end{center}
\caption{\label{tab:chinaCilin}The results as obtained using the $C^{r}_i$ estimates
  of \cite{Jiang:2009uf}. Both the fits include the same observables as the fit
  in Table~\ref{tab:fitalllin}, i.e. as fit All but with the linear fit from NA48/2. 
  In the second column the coefficient $\alpha$
  multiplied by the $C^{r}_i$ has been included as fitting parameter. The $L^r_i$
  are given at $\mu=0.77$ GeV.}
\end{table}
\begin{table}
\begin{center}
\begin{tabular}{|c|ccc|}
\hline
 & $p^{2}$&$ p^{4}$& $p^{6}$\\
\hline
$m^{2}_{\pi}$ & $0.624$ & $0.384$ & $-0.008$\\
$m^{2}_{K}$   & $0.667$ & $0.189$ & $0.144$\\
$m^{2}_{\eta}$& $0.716$ & $0.149$& $0.135$\\
$F_\pi/F_0$       &$1$ & $0.354$ & $0.142$\\
$F_K/F_0$         &$1$ & $0.531$ & $0.225$\\
$F_K/F_\pi$   &$1$ & $0.177$& $-0.020$ \\
\hline
\end{tabular}
\end{center}
\caption{\label{tab:chinaCilinb}The convergence for
  the fit where $\alpha$ is left free is shown. The normalizations are the
  same as explained in Table~\ref{tab:fit10b}. Now $F_0=0.062$ GeV.
  Fit as in rightmost column of Table~\ref{tab:chinaCilin}.}
\end{table}

In Table~\ref{tab:chinaCilin} and \ref{tab:chinaCilinb} 
we quote the results obtained with the $C^r_i$
of \cite{Jiang:2009uf}, but fitting the slope of the linear fit for the $F_s$
formfactor as in the fit of Table~\ref{tab:fitalllin}. Conclusions similar to
the ones drawn for Table~\ref{tab:chinaCi} hold here too.

\subsection{Comparison with a recent $L_5^r$ determination}
\label{sect:comparison}

The authors of \cite{Ecker:2010nc} propose a simplification of the NNLO
predictions of ChPT to perform fits to lattice data points. As an example
 they fit the lattice results \cite{Durr:2010hr} for $F_{K}/F_{\pi}$ with
 their approximation to the two-loop ChPT prediction. Since too many
 couplings appear at NNLO they are forced to fit only a few of them and
 fix the others. They decide to fit $L^{r}_{5}$ and the only
 two $C^{r}_{i}$-combinations contributing: $C^{r}_{14}+C^{r}_{15}$ 
and $C^{r}_{15}+2C^{r}_{17}$. $C^{r}_{15}$ is $1/N_{c}$ suppressed and therefore
 set to $0$ in the resonance saturation model. $C^{r}_{14}$ and $C^{r}_{17}$
 also do not get contributions from the resonances we included. In other
 models \cite{Ecker:2010nc} $C^{r}_{14}$ and $C^{r}_{17}$ are estimated
 to be negative, but are very small in absolute value. In 
\cite{Ecker:2010nc} the other $L^{r}_{i}$, appearing at NNLO in $F_{K}/F_{\pi}$,
 are set to the values of fit 10 \cite{Amoros:2001cp}.
The results of the fit to lattice data are quoted in the first column of 
Table \ref{tab:comparison}.  For comparison we quote in the same table
 the values as obtained by our best fits, including those for the random $C_i^r$
search described in Section~\ref{sect:randomCi} and quoted in Table~\ref{tab:Ci}
in the appendix.
\begin{table}
\begin{center}
\begin{tabular}{|c||c|c|c|c|}
\hline
    & \cite{Ecker:2010nc} & fit All & best reso & best rand \\
\hline
$10^3 L^r_5$        & $0.76\pm0.09$ &$0.58\pm0.13$& $1.40\pm0.009$ & $1.40\pm0.009$\\
$10^5 C^{r}_{14}+C^{r}_{15}$ & $0.31\pm0.07$ &$0$          & $-0.99$ & $-1.06$\\
$10^5 C^{r}_{15}+2C^{r}_{17}$& $1.10\pm0.14$   &$0$          & $0.02  $ & $2.01$\\
\hline
\end{tabular}
\caption{\label{tab:comparison} The values of the
 couplings $L^{r}_{5}$, $C^{r}_{14}+C^{r}_{15}$ and $C^{r}_{15}+2C^{r}_{17}$ as
 obtained from a fit to lattice data \cite{Ecker:2010nc} (second column) and
from our fits. Notice that the resonance model used for fit All estimates
 the combinations of $C^{r}_{i}$ occurring here as zero. The columns best
 reso and best rand are taken from Tables~\ref{tab:random} and \ref{tab:Ci}
 and are discussed in Section~\ref{sect:randomCi}.
All the values are at the
 scale $\mu=0.77$~GeV.}
\end{center}
\end{table}
The table shows that the value of $L^{r}_{5}$ in fit All is compatible with
the result of \cite{Ecker:2010nc} within uncertanties and even more
compatible when we look at the fit $F_K/F_\pi$ in Table~\ref{tab:fitall}
where we required $L_4^r=0$ as in \cite{Ecker:2010nc}.
The fits best reso and best rand
instead are very different. For these last two fits we can also compare
the values of the
$C^{r}_{i}$-combinations. As far as regards the combination
$C^{r}_{14}+C^{r}_{15}$ our random sets acquire both a negative value in
contrast with \cite{Ecker:2010nc}. The second combination instead agrees upon
the sign, but are rather different in value.

For further comparison we
can do also a fit similar to fit All but setting in addition
$10^{5}C^{r}_{14}=0.31$ and  $10^{5}C^{r}_{17}=0.55$. The results are given in
Table~\ref{tab:fitEMN}.
\begin{table}
\begin{center}
\begin{tabular}{|c|c|}
\hline
             & $C_i^r$\cite{Ecker:2010nc}  \\
\hline
$10^3 L_1^r$ &  $0.88\pm0.09$   \\
$10^3 L_2^r$ &  $0.53\pm0.21$     \\
$10^3 L_3^r$ &  $-2.97\pm0.43$   \\
$10^3 L_4^r$ &  $0.89\pm0.83$   \\
$10^3 L_5^r$ &  $0.30\pm0.09$   \\
$10^3 L_6^r$ &  $0.39\pm0.97$   \\
$10^3 L_7^r$ &  $-0.02\pm0.16$   \\
$10^3 L_8^r$ &  $0.13\pm0.20$   \\
\hline
$\chi^2$ (dof)    & $2.03$ $(4)$      \\
\hline
\end{tabular}
\end{center}
\caption{\label{tab:fitEMN}The fit  as obtained using the same input as fit
  All and the $C^{r}_i$ as in the resonance estimate of Table~\ref{tab:Ci}
  with the exception of $C^{r}_{14}$  and
  $C^{r}_{17}$ set respectively to $0.31\times10^{-5}$ and $0.55\times10^{-5}$.
  The $L^r_i$ are given at $\mu=0.77$ GeV. }
\end{table}

The fit obtained is different from fit All, in particular we find an even lower
$L_5^r$ and the $\chi^{2}$ is somewhat larger. 
The value of $L^{r}_{5}$ has decreased quite a bit while
the others are compatible with fit All but the
central values for $L_4^r$ and $L_6^r$ are rather large.
Finally notice that the $L^{r}_{i}$ of fit All are very different from the
ones of fit 10 and might therefore affect the findings of \cite{Ecker:2010nc}.
We cannot draw any further conclusions at present. In future work we intend
to include more lattice results which should clarify this issue.

\section{Releasing the $C_i^r$}
\label{sect:randomCi}

All the fits presented in the previous section have unusual NNLO corrections to
the masses and many also to the decay constants. In addition, if we included the
requirement that the $\bar\ell_i$ were also fitted well we could not find
a simple good fit.

An additional reason to go beyond what we have is that all the estimates
used above with the exception of the singlet $\eta$ contribution
only contribute to the NNLO LECs\footnote{This is true with the exception of
terms involving $\langle\chi_-\rangle$ which can get produced by the equations
of motion.}
 that are leading in $N_c$.
In the masses and decay constants
in addition the estimates from the resonance exchange give no contribution
at leading order in $N_c$ at all.
This is an unsatisfying situation, we do expect that the masses and decay constants
should get some contribution from the NNLO constants. Inspection of
the relations between the $\bar{\ell}_i$ and the $SU(3)$ LECs \cite{Gasser:2007sg}
shows that only combinations of the $C_i^r$ appear that are suppressed by $N_c$.
Thus especially the problem with $\bar\ell_2$ above requires some nonzero values
for the $N_c$ suppressed constants.

\begin{figure}
\begin{center}
\includegraphics{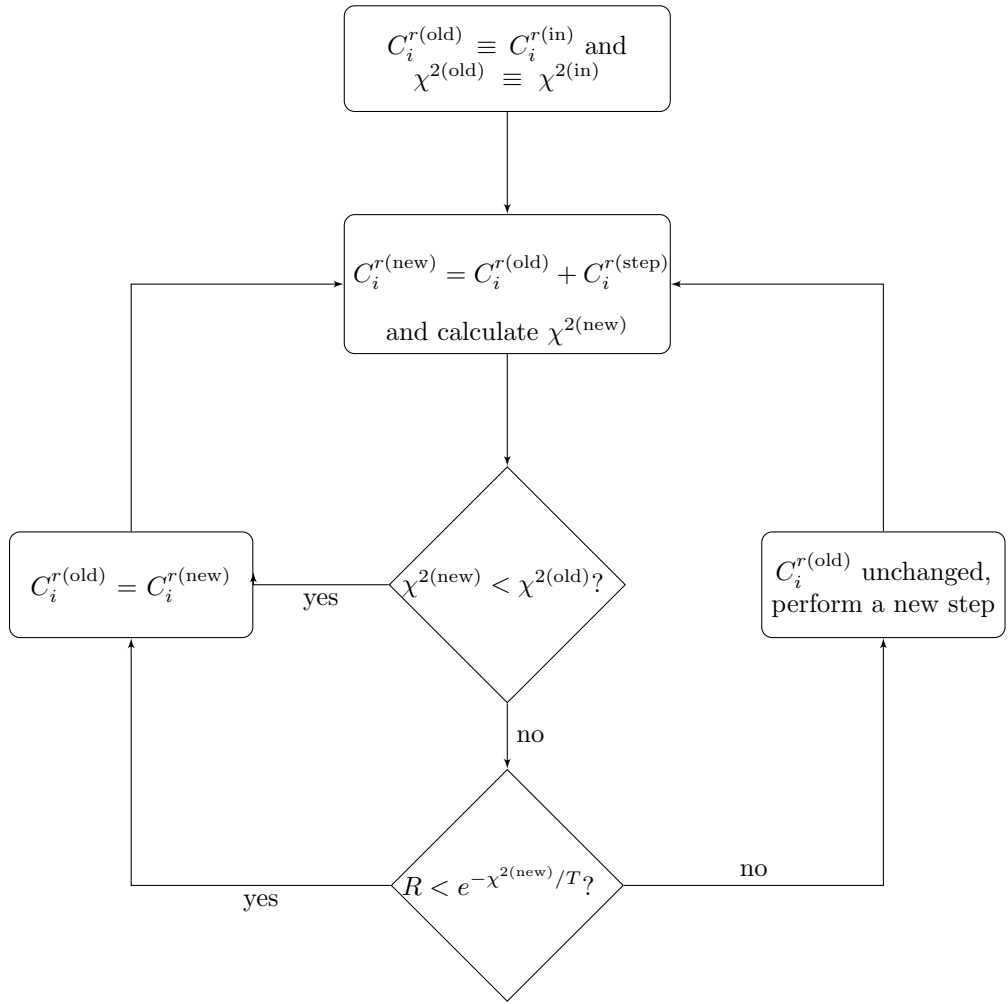}
\end{center}
\caption{\label{fig:flow}Algorithm used to select the random $C^r_i$. 
It has been started with different values of the initial
$C^{r(\rm{in})}_i$, as explained in the text. In the
bottom decision square $R$ is a random number selected with a uniform
distribution in the interval (0,1), while $T$ is a parameter set such that
it is of the same order of magnitude of the $\chi^2$. More details can be found
in the text.}
\end{figure}

We could in principle allow all the $C_i^r$ to be free and include them in the fit
as well. However, from our earlier work in \cite{Bijnens:2009zd} it is clear that
with the inputs used at present there are enough free combinations of the $C_i^r$
to fit all physical inputs directly.
For this reason we also have explored another technique of $C^{r}_i$
estimate, based on a random walk\footnote{The idea was born thanks to a
discussion with Juerg Gasser and Gerhard Ecker.} method. Hereafter we
describe the main features of the algorithm used. See also the flowchart in
Figure~\ref{fig:flow}. The algorithm is a version of simulated annealing.

We first start with an initial set  $C^{r(\rm{in})}_i=C^{r({\rm old})}_i$. 
These are chosen to be
\begin{enumerate}
\item random (with a
size given by $1/3/(16\pi^2)^2$ for those leading in $1/N_c$ 
and $1/3$ of that for the subleading ones),
\item all zero
\item as obtained by
resonance estimate (see Table~\ref{tab:Ci})
\item as obtained by multiplying the constants of \cite{Jiang:2009uf} by
  $0.27$ (see Table~\ref{tab:Ci}).
\end{enumerate}
Then we perform the fit on the $L^r_i$ using those
$C^{r({\rm old})}_i$.
After this we take a random step according to the formula
\be\label{randstep}
C^{r({\rm new})}_i=C^{r({\rm old})}_i+C^{r({\rm step})}_i\equiv C^{r({\rm old})}_i+
\frac{1}{(16\pi^2)^2}\epsilon r_i\,,
\ee
where $r_i$ in (\ref{randstep}) is a random number generated through a uniform
distribution in the interval $(-1,1)$
and we have used $\epsilon=0.01$ and $0.001$.
For those $C^r_i$ that are $N_c$
suppressed we further multiply $C^{r({\rm step})}_i$ by $1/3$. In this way the
random generated $C^r_i$ set still respects the large-$N_c$ suppressions
to a certain extent.  We perform the fit of the $L^r_i$
using the new set
$C^{r({\rm new})}_i$ and we check the $\chi^2$ obtained.
If the $\chi^2$ decreases then we substitute the $C^{r({\rm old})}_i$ with
$C^{r({\rm new})}_i$. Sometimes we also allow $C^{r({\rm new})}_i$ to be selected
even though the corresponding $\chi^2$ is not smaller than the previous
one (see last step in the flow diagram of Figure~\ref{fig:flow}). This is done
so to let the $C^{r}_i$ to take quite different values and thus to
test as many different sets as possible.
It is also needed for our algorithm to be able to move out of local minima.
We find that, when we let our algorithm run long enough,
we cover quite many different sets of $C^r_i$. 
We chose different
initial $C^r_i$ to widen their range of variability. In addition we have
started the random walk from the same starting point several times including
different random starting points.
We chose as random starting points $1/3/(16\pi^2)^2$ with the extra factor of $1/3$
since without the extra $1/3$ we never reached a $\chi^2$ smaller than one.

The fits are performed including the same input as for fit All but with
a few extra requirements.
We add as input the curvature of the scalar formfactor $c^\pi_S$ in
(\ref{sff}) and all the $\bar{\ell_i}$ of the last column of Table~\ref{tab:li}.
We do not instead demand to fit all the $\pi\pi$ and $\pi K$ scattering
parameters, since it costs in terms of computing time. So, as done for fit
All, we include only $a^0_0$, $a^0_2$, $a^{1/2}_0$ and $a^{3/2}_0$. Notice
that we do not find any large discrepancies when we add more scattering
parameters in fit All, as was remarked at the end of
Section~\ref{sect:newfits}.

We also require a good convergence of the masses and decay constants
expansions. The reason is that in this way we have the possibility to
``select'' those $C^r_i$ granting us convergence for those
quantities. This also allows to keep under control the quality of the
fits. Otherwise too much freedom would be left to the $C^r_i$
constants, and many different fits with a low $\chi^2$ but with very bad
convergences, can be reached. Clearly such
convergence constraints have a strong effect on the $L^r_i$ constants
as described in the next Section~\ref{sect:conv}. 

\subsection{Convergence constraints}
\label{sect:conv}

We devote this section to a discussion of the convergence constraints imposed on the
masses and decay constants in the fits with random $C^r_i$. Let us
first show the case of the decay constants. 

We performed the fits constraining the NNLO
contributions to $F_\pi$, $F_K$ and $F_K/F_\pi$ constants to be small,
i.e. less than the $10\%$ of the LO ones. 
Remember that the expansions for the $F_\pi$ and $F_K$ decay constants are
\ba
F_K&=&F_0+\left.F_K\right|_{p^4}+\left.F_K\right|_{p^6},\nonumber\\
F_\pi&=&F_0+\left.F_\pi\right|_{p^4}+\left.F_\pi\right|_{p^6},\nonumber\\
\ea
and that in our fits, as explained in Section~\ref{sect:massesanddecay}, we
include their ratio as  
\ba
\frac{F_K}{F_\pi}\approx 1+
\underbrace{
\left.\frac{F_K}{F_0}\right|_{p^4}-\left.\frac{F_{\pi}}{F_0}\right|_{p^4}}
_{\rm NLO}
\underbrace{
+\left.\frac{F_K}{F_0}\right|_{p^6}-\left.\frac{F_\pi}{F_0}\right|_{p^6}
-\left.\frac{F_K}{F_0}\right|_{p^4} \left.\frac{F_{\pi}}{F_0}\right|_{p^4}
+\left.\frac{F_\pi}{F_0}\right|^2_{p^4}}
_{\rm NNLO}.
\ea
Specifically the convergence constraints are included through the
following partial $\chi^2_{i(\rm{part})}$:
\ba\label{Fconstr}
&&\chi^2_{i(\rm{part})}=\left(\left.\frac{F_\pi}{F_0}\right|_{p^6}/0.05\right)^2,\qquad\chi^2_{i(\rm{part})}=\left(\left.\frac{F_K}{F_0}\right|_{p^6}/0.05\right)^2,\nonumber \\
&&\chi^2_{i(\rm{part})}=\left(\left.\frac{F_K}{F_\pi}\right|_{\rm NNLO}/0.07\right)^2,
\ea
With (\ref{Fconstr}) we have been able to find many fits with a good
convergence. On the other hand the NLO
contribution for $F_\pi$ turns out to be smaller than the expected $30\%$.
The reason resides in the third of the
relations in (\ref{Fconstr}).  Requiring all the NNLO pieces for all the decay
constants to be small implies that the single contributions $F_\pi|_{p^6}$ and $F_K|_{p^6}$ are small. But also
the term $(F_\pi/F_0)|^2_{p^4}$ must be small, otherwise
the NNLO contribution of $F_K/F_\pi$ is allowed to be large. This
leads to small NLO corrections for $F_\pi$ and thus a $F_0\approx F_\pi$.

We apply similar restrictions also to the masses
\ba\label{Mconstr}
 \chi^2_{i(\rm{part})}=\left(\left.\frac{m^2_M}{m^2_{M\,0}}\right|_{p^6}/0.1\right)^2
\ea
where $M$ stands for $\pi$, $K$ and $\eta$ mesons and $m_{M\,0}$ are the
leading order contributions to the masses. We have kept the value
of $m_s/\hat m$ fixed at 27.8 as for fit All.

We conclude this section with a final remark. One might wonder why we have not
imposed similar constraints also for fit All, since these could improve the
convergence of the expansions. The reason is that when we require
them, we obtain a reasonably good fit ($\chi^2\approx8$ with 10 degrees of
freedom ) and  with better constrained $L^r_4$ and $L^r_6$. But it also
causes much worse predictions for the $\bar{\ell_i}$, e.g. $\bar{\ell_3}\approx
6.5$.

\subsection{Results}
\label{sect:randres}

\begin{figure}
\begin{center}
\includegraphics[angle=270,width=0.8\textwidth]{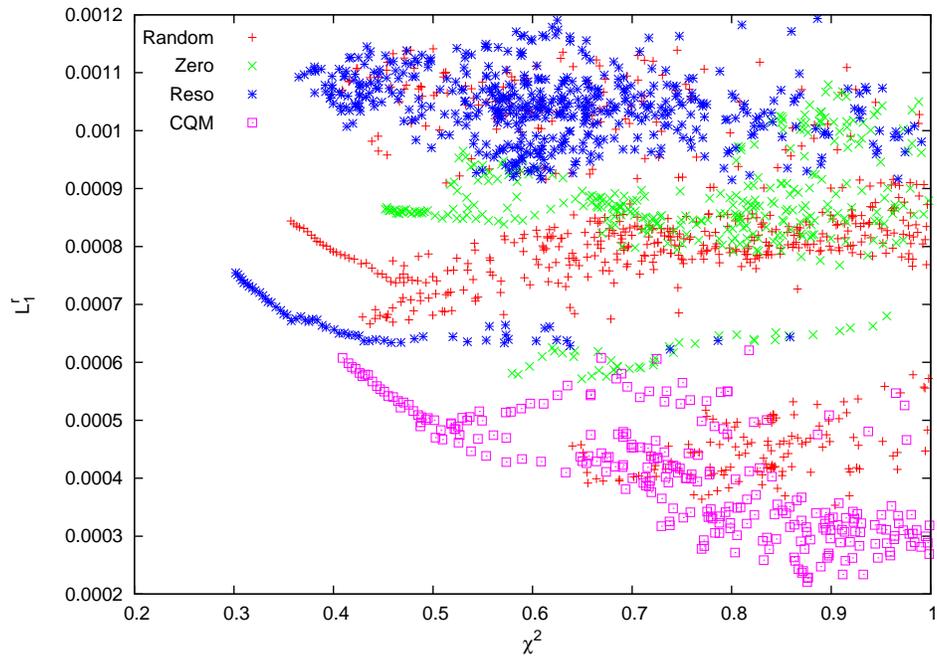}
\end{center}
\caption{\label{fig:L1} The values of $L^r_1$ for the
  random walk fits of Table~\ref{tab:ranges}. In the plot all the fits are
  collected. The ranges
  of variability for $L_1^r$ is quite large. In the picture it
  is also evident how the two values of the couplings depend somwehat on the
  different initial values of the $C^r_i$ constants. The $L^r_i$ are at
  the scale $\mu=0.77$ GeV.}
\end{figure}

Now we are ready to show the outcomes of our studies when the $C^r_i$ are set
to random values using the procedure of Figure~\ref{fig:flow} and with the
constraints listed in Section~\ref{sect:conv}
above. First of all we must point out that due to the freedom we allow to the
$C^r_i$ many different fits of the $L^r_i$ have a low $\chi^2$. We set initial
$C^r_i$ equal to zero, resonance exchange or chiral quark model estimates
as well more random starts.
We easily reach $\chi^2<1$, and we found many fits with $\chi^2\approx
0.5$. Reducing the steps of the random walk we can even find smaller
values. However once the $\chi^2$ reaches a reasonably low value, e.g. $1$,
there is no apparent reason why one should prefer one specific fit to
another. Due to the several different sets of $C^r_i$ under study,
we can only quote the ranges where the $L^r_i$ vary and where we obtain a
$\chi^2<1$. 
Such ranges are quoted in Table~\ref{tab:ranges} and are obtained, starting from
different initial sets $C^{r({\rm in})}_i$.
Keep in mind that these ranges depend on the $C^r_i$ chosen. 
Plugging in a $L^r_i$ fit without the corresponding $C^r_i$
will not produce any sensible results. The way we determined those numbers is shown
in Figure~\ref{fig:L1} on the example for $L_1^r$ where we have plotted
a number of fits that gave $\chi^2<1$ for the different starting points.
We have typically stopped the fits when a $\chi^2$ of about 0.4 or below was
found and the tails at low $\chi^2$ are an artefact, they were done with
runs with a very low $\epsilon$ and a very low $T$.
\begin{table}
\begin{center}
\begin{tabular}{|l|c c c c |}
\hline
$C^{r({\rm in})}_i$ & resonance       & zero           & CQM            &random\\
\hline                                                                  
$10^3 L_1^r$       & $(0.6,1.2)$     &  $(0.5,1.1)$    & $(0.2,0.6)$  &$(0.3,1.2)$\\
$10^3 L_2^r$       & $(0.4,1.2)$    &   $(0.2,0.8)$  & $(-0.2,0.5)$   &$(0,1.6)$\\
$10^3 L_3^r$       & $(-5.5,-3.0)$ &  $(-4.5,-3.0)$  & $(-3.3,-1.2)$&$(-6.2,-2.0)$ \\
$10^3 L_4^r$       & $(0.15,0.35)$   &  $(-0.1,0.25)$ & $(-0.1,0.2)$&$(0.05,0.35)$ \\
$10^3 L_5^r$       & $(1.2,1.6)$   &  $(1.35,1.5)$  & $(1.25,1.55)$  &$(1.2,1.6)$ \\
$10^3 L_6^r$       & $(-0.05,0.25)$  &  $(-0.2,0.2)$&$(-0.15,0.15)$ &$(-0.05,0.3)$ \\
$10^3 L_7^r$       & $(-0.45,-0.1)$ &$(-0.43,-0.18)$&$(-0.43,-0.24)$&$(-0.45-0.2)$ \\
$10^3 L_8^r$       & $(0.4,0.8)$    &  $(0.45,0.72)$  & $(0.5,0.731)$  &$(0.45,75)$\\
\hline
\end{tabular}
\caption{\label{tab:ranges} The ranges for $L^r_i$ values as obtained
  changing the $C^r_i$ according to a random walk algorithm. The different
  ranges correspond to different initial values for the $C^r_i$. With CQM we
  indicate the $C^r_i$ of the chiral quark model. All the fits
  have a $\chi^2<1$. For all
  the fits $\mu=0.77$ GeV.
  See the
  description in the text for further details on how the fits have been performed.}
\end{center}
\end{table}

The results are rather cumbersome. $L^r_1$, $L^r_2$
and $L^r_3$ look quite free to vary in large ranges. Notice also that
the large $N_c$ relation $2L^r_1\approx L^r_2$ is still not recovered. On the
other hand, due to the converge constraints of Section~\ref{sect:conv} we
narrow the intervals for the other constants. Especially $L^r_5$ takes
a large value and $L^r_4$ a small one. 
As
explained in Section~\ref{sect:conv}, we essentially require
that $F_K/F_\pi$, $F_K$ and $F_\pi$ have small NNLO corrections
which give that $F_K/F_\pi$ is given by the NLO and thus determines
$L^r_5$ at a fairly large value. That $F_\pi$ has small corrections at all then in
turn requires a fairly small $L_4^r$.
The choice to constrain the
convergence of those quantities is dictated by the lack of information to
constrain more the $C^r_i$. When we release such constraints indeed we find
different looking fits, but affected by a bad convergence.
Somewhat more surprising is that the $L_6^r$ typically takes on values that are
smaller than $L_8^r$.

As you can see no clear final conclusion can be drawn with such results. 
When we performed this study we were hoping not to find as many good fits
and smaller ranges for the $L_i^r$. The study
shows instead that it is very difficult, if not impossible, to narrow the
ranges for the $L^r_i$
with such a poor knowledge of the $C^r_i$. On the other hand it also shows
that fits of the $L^r_i$ with good convergence do exist, if the $C^r_i$ are
changed.
\begin{table}
\begin{center}
\begin{tabular}{|c|cc|}
\hline
        $C^{r}_i$ & best reso & best random \\
\hline
$10^3 L_1^r$ &  $0.75\pm0.09$    &$0.85\pm0.09$ \\
$10^3 L_2^r$ &  $0.81\pm0.45$    &$0.54\pm0.05$   \\
$10^3 L_3^r$ &  $-3.91\pm0.28$   &$-3.51\pm0.28$ \\
$10^3 L_4^r$ &  $0.16\pm0.10$    &$0.20\pm0.10$ \\
$10^3 L_5^r$ &  $1.40\pm0.09$    &$1.40\pm0.09$ \\
$10^3 L_6^r$ &  $0.10\pm0.14$    &$0.12\pm0.14$ \\
$10^3 L_7^r$ &  $-0.32\pm0.13$   &$-0.32\pm0.13$ \\
$10^3 L_8^r$ &  $0.64\pm0.16$    &$0.63\pm0.16$ \\
\hline
$\chi^2$    & $0.30$    & $0.36$  \\
\hline
\end{tabular}
\end{center}
\caption{\label{tab:random}The results as obtained using the $C^{r}_i$ 
  from the best $\chi^2$ found starting from the resonance
or from a completely random one as described in the text. The $L^r_i$
  are given at $\mu=0.77$ GeV. The corresponding $C^r_i$ sets can be found in
  Table~\ref{tab:Ci} in appendix.}
\end{table}
\begin{table}
\begin{center}
\begin{tabular}{|c|ccc|ccc|}
\hline
$C_i^r$ & \multicolumn{3}{c|}{best reso} & \multicolumn{3}{c|}{best random}\\
\hline
 & $p^{2}$&$ p^{4}$& $p^{6}$& $p^{2}$&$ p^{4}$& $p^{6}$\\
\hline
$m^{2}_{\pi}$ & $0.987$ & $0.021$ & $-0.008$ & $0.993$ & $0.021$ & $-0.012$ \\
$m^{2}_{K}$   & $1.057$ & $-0.054$ & $-0.003$& $1.060$ & $-0.058$ & $-0.002$\\
$m^{2}_{\eta}$& $1.132$ & $-0.133$& $0.001$  & $1.136$ & $-0.135$& $-0.001$  \\
$F_\pi/F_0$       &$1$ & $0.178$ & $-0.010$      &$1$ & $0.187$ & $-0.010$  \\
$F_K/F_0$         &$1$ & $0.395$ & $0.009$       &$1$ & $0.404$ & $0.011$   \\
$F_K/F_\pi$   &$1$ & $0.217$& $-0.020$       &$1$ & $0.217$& $-0.020$       \\
\hline
\end{tabular}
\end{center}
\caption{\label{tab:randomb}The convergence for
  the best $\chi^2$ found starting from the resonance estimate, here $F_0=0.079$ GeV.
  Fit as in left column of Table~\ref{tab:random}.
  The convergence for
  the best $\chi^2$ found starting from the fully random estimate,
   here $F_0=0.078$ GeV.
  Fit as in left column of Table~\ref{tab:random}.
}
\end{table}
In Table~\ref{tab:random} we show the $L_i^r$ obtained for the smallest
$\chi^2$ found starting from the resonance estimate and from fully random
$C_i^r$ as described above and we show the convergence for some quantities
in those two fits in Table~\ref{tab:randomb}.
The fits with very low $\chi^2$ we have obtained, such as the two shown here,
tend to have similar expansions for the masses and the decay constants.
In order to see how the various $C_i^r$ look like we have added the values for
these two fits in the appendix. 

We can draw some conclusions by studying correlations. The effect of
$\bar\ell_2$ is very visible if we plot for the various fits with $\chi^2<1$
$L_2^r$, which is the NLO dependence of $\bar\ell_2$
on the LECs, versus $2C^r_{13}-C^r_{11}$, which is the dependence
on the NNLO LECs. The tight correlation shows that
the NNLO contribution depends very little on the other $L^r_i$.
More precisely, this shows the NLO LEC combination at NLO versus the NNLO LEC combination
at NNLO and that the NNLO contribution depends fairly little on the value of the $L_i^r$.
\begin{figure}
\begin{center}
\includegraphics[angle=270,width=0.8\textwidth]{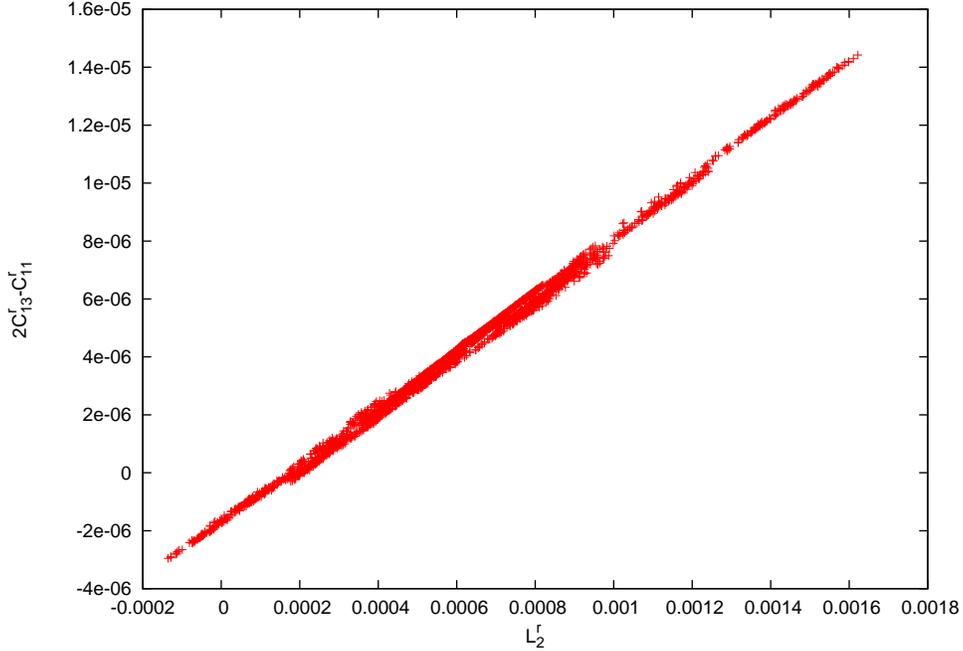}
\end{center}
\caption{\label{fig:l2bar}
The correlations between $L_2^r$ and $2C^r_{13}-C^r_{11}$.
This results from including the value of $\bar\ell_2$ in the fit.}
\end{figure}
We get more of those constraints directly. A very similar one is from the
value of $\bar\ell_1$ shown in Figure~\ref{fig:l1bar}.
\begin{figure}
\begin{center}
\includegraphics[angle=270,width=0.8\textwidth]{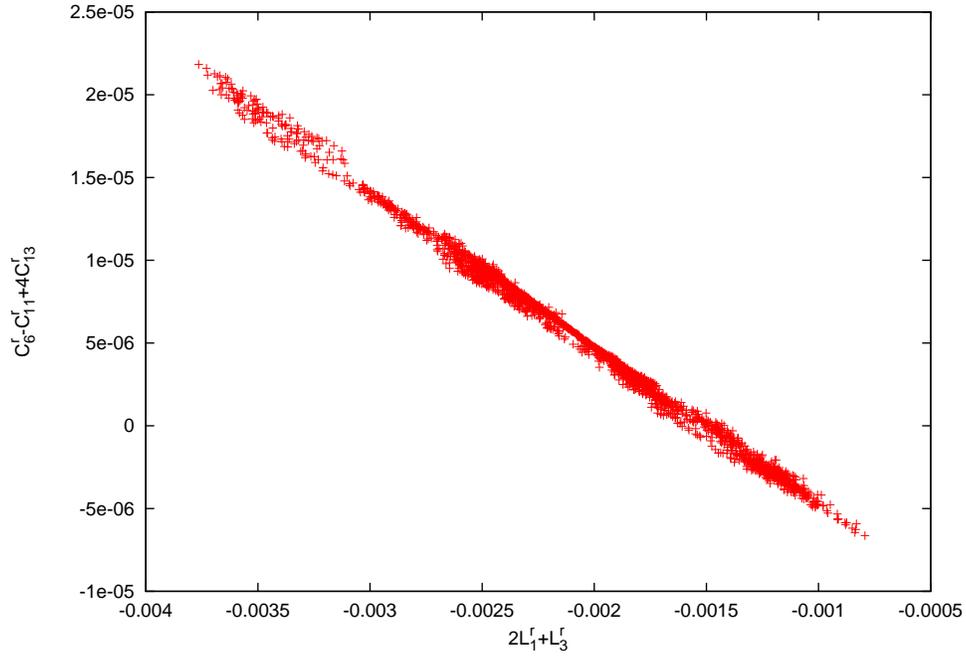}
\end{center}
\caption{\label{fig:l1bar}
The correlations between $2L_1^r+L_3^r$ and $C^r_6-C^r_{11}+4C^r_{13}$.
This results from including the value of $\bar\ell_1$ in the fit.}
\end{figure}
The correlations for other observables tend to be weaker indicating that
the NNLO contributions are more dependent on the value of the $L_i^r$
for these cases. We show examples with a weaker but still existing correlation
in Figure~\ref{fig:fkfpi} where the correlations resulting from
$F_K/F_\pi$ are shown and in Figure~\ref{fig:scalarradius} for
$\langle r^2\rangle^\pi_S$. In both cases we have plotted on the horizontal axis
the combination of $L_i^r$ the quantity depends on at NLO and
on the vertical axis the combination of $C_i^r$ the quantity depends on at NNLO.
\begin{figure}
\begin{center}
\includegraphics[angle=270,width=0.8\textwidth]{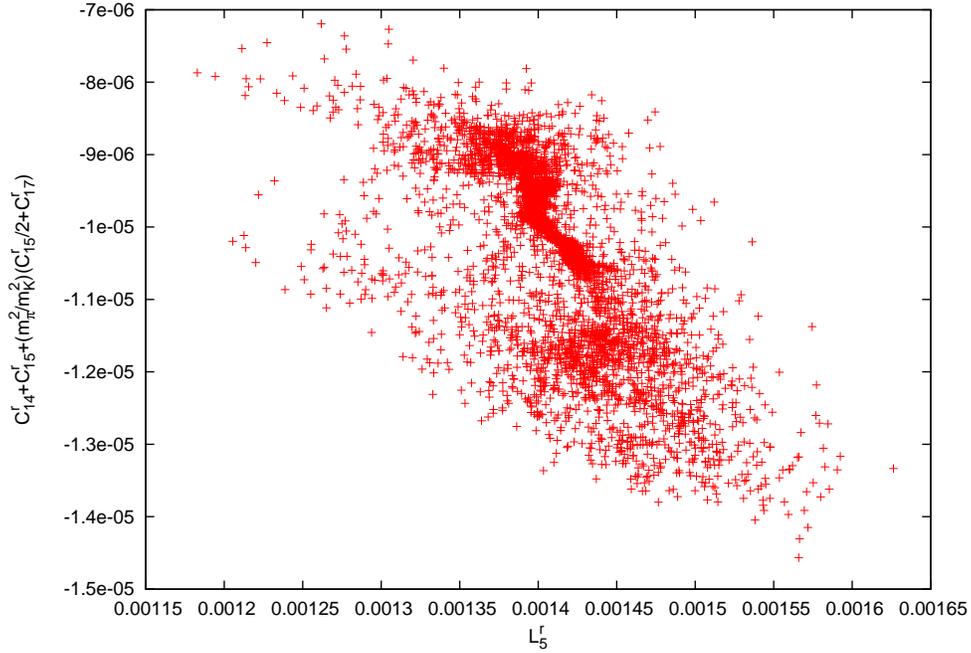}
\end{center}
\caption{\label{fig:fkfpi}
The correlations resulting from $F_K/F_\pi$.}
\end{figure}
\begin{figure}
\begin{center}
\includegraphics[angle=270,width=0.8\textwidth]{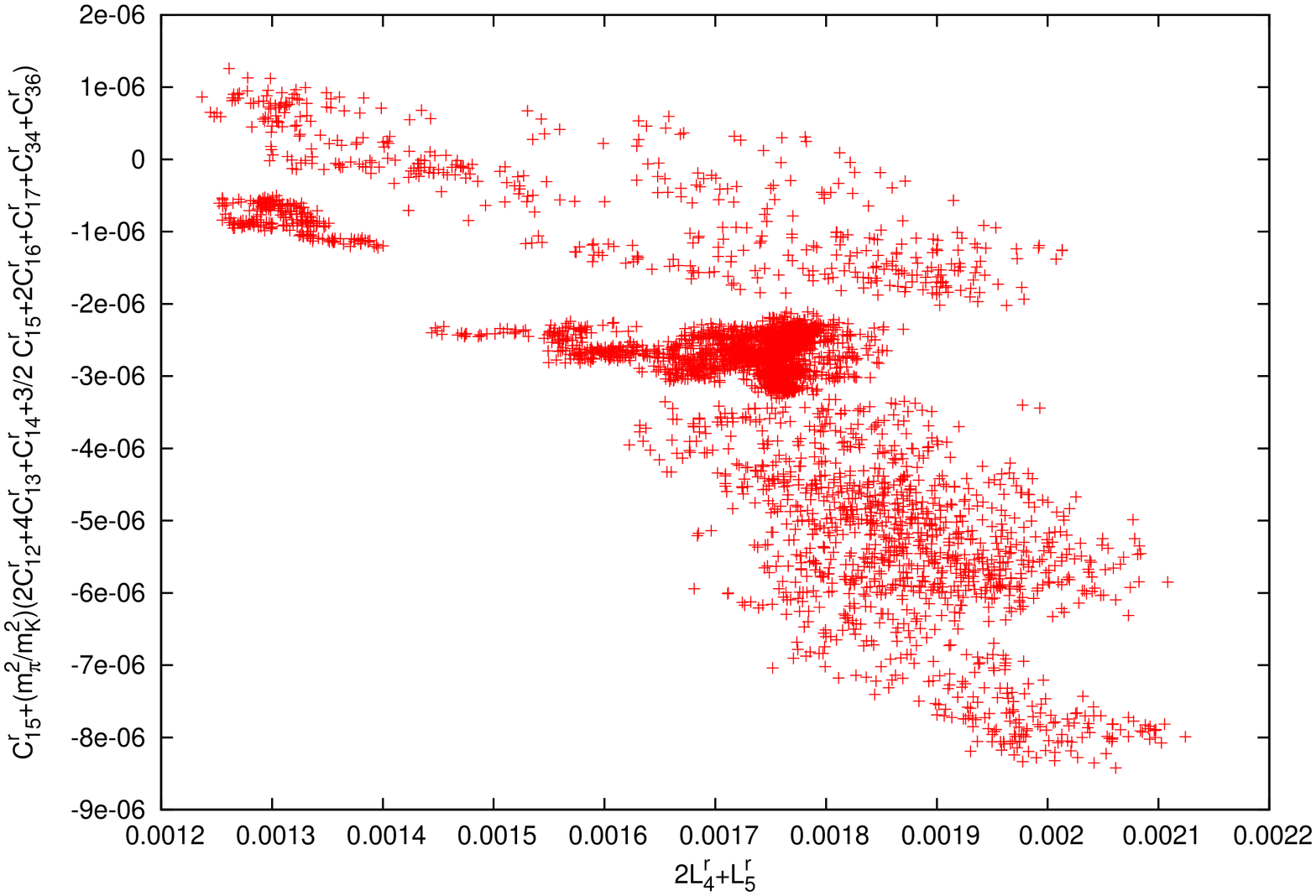}
\end{center}
\caption{\label{fig:scalarradius}
The correlations resulting from the pion scalar radius $\langle r^2\rangle^\pi_S$.}
\end{figure}

\begin{figure}
\begin{center}
\includegraphics[angle=270,width=0.8\textwidth]{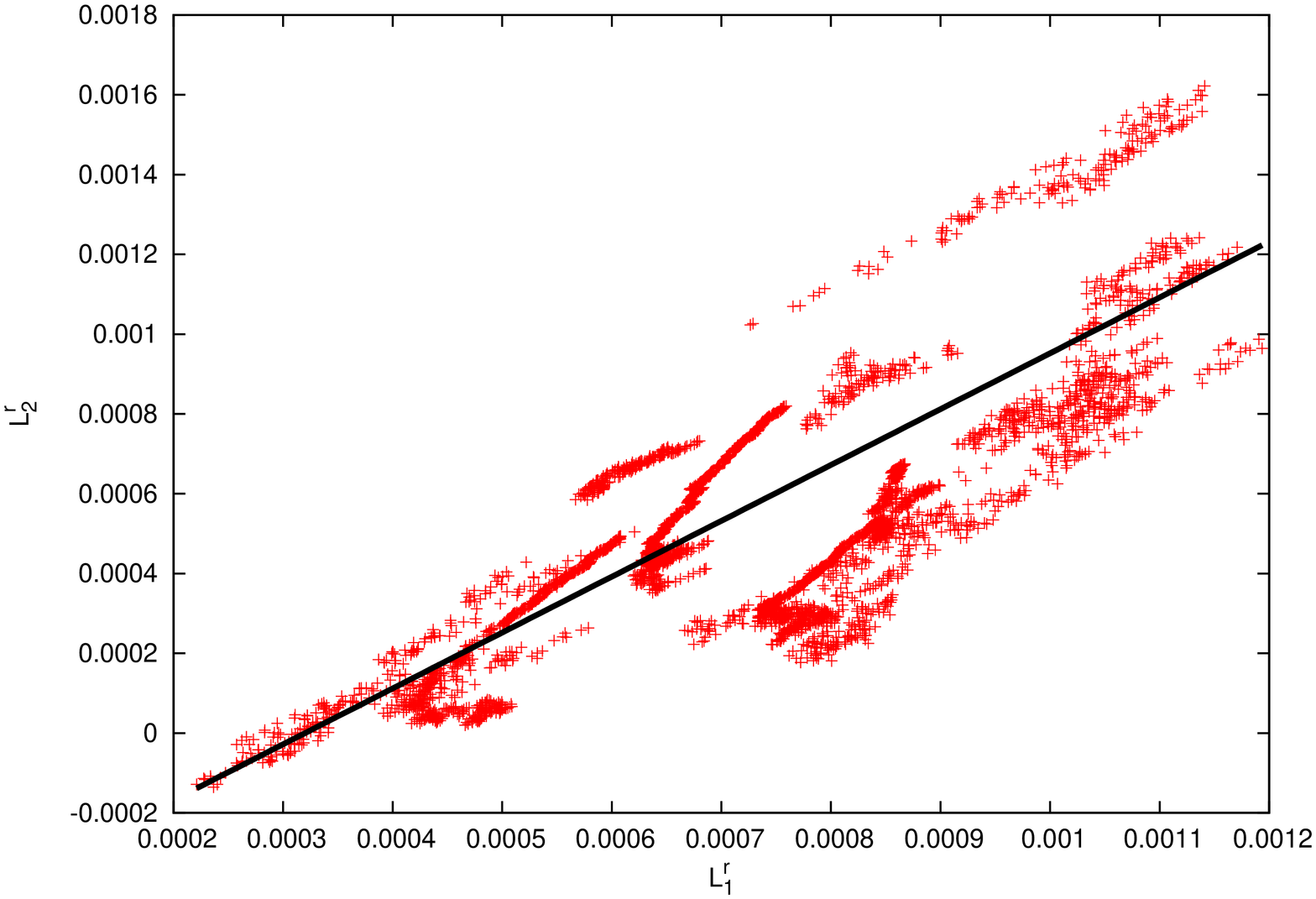}
\end{center}
\caption{\label{fig:L12}
The correlations between $L_1^r$ and $L_2^r$,
the curve shown is $L_2^r = 1.4(L_1^r-0.00032)$.}
\end{figure}

\begin{figure}
\begin{center}
\includegraphics[angle=270,width=0.8\textwidth]{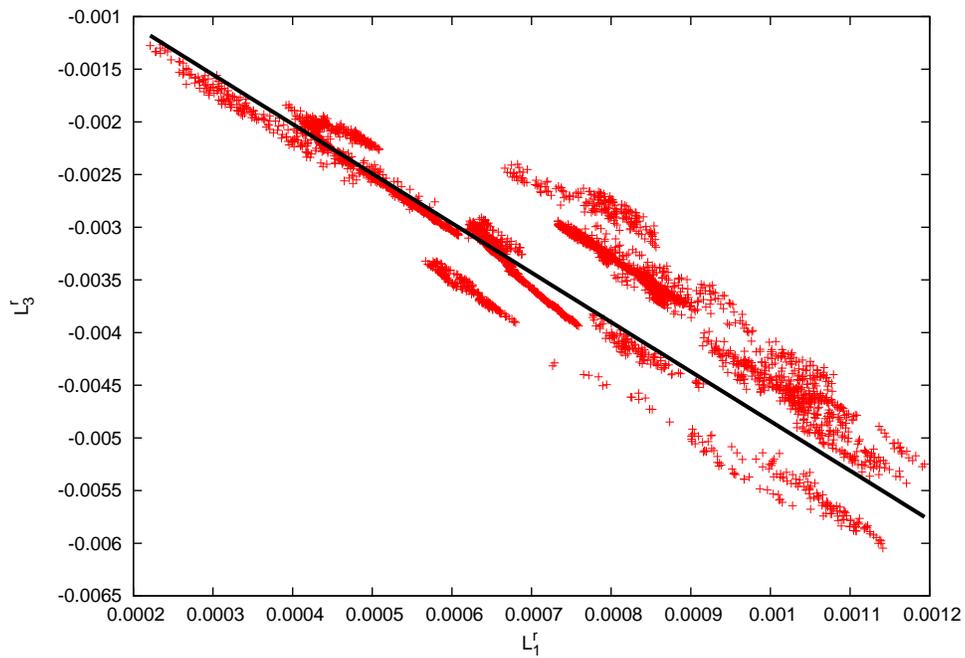}
\end{center}
\caption{\label{fig:L13}
The correlations between $L_1^r$ and $L_3^r$,
the curve shown is $L_3^r = -4.7(L_1^r+0.00003)$.}
\end{figure}

\begin{figure}
\begin{center}
\includegraphics[angle=270,width=0.8\textwidth]{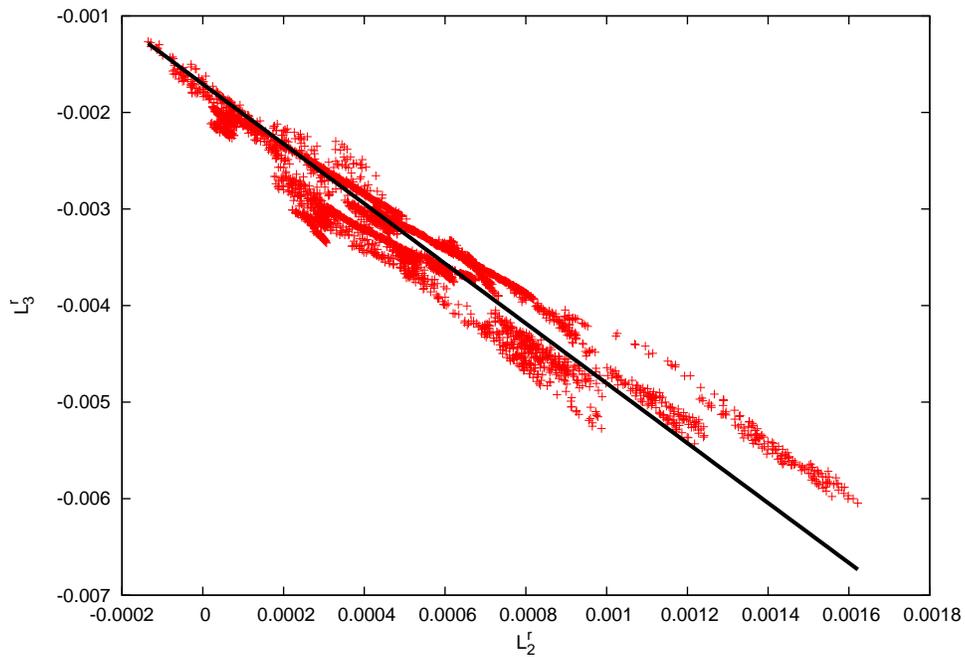}
\end{center}
\caption{\label{fig:L23}
The correlations between $L_2^r$ and $L_3^r$,
the curve shown is $L_3^r = -3.1(L_2^r+0.00055)$.}
\end{figure}

\begin{figure}
\begin{center}
\includegraphics[angle=270,width=0.8\textwidth]{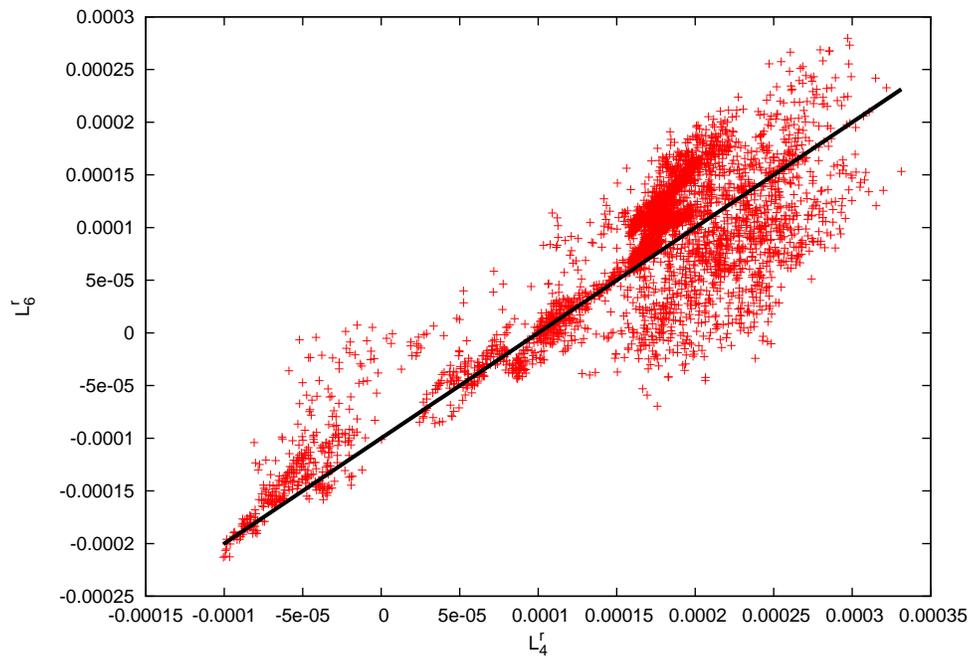}
\end{center}
\caption{\label{fig:L46}
The correlations between $L_4^r$ and $L_6^r$,
the curve shown is $L_6^r = L_4^r-0.0001$.}
\end{figure}

\begin{figure}
\begin{center}
\includegraphics[angle=270,width=0.8\textwidth]{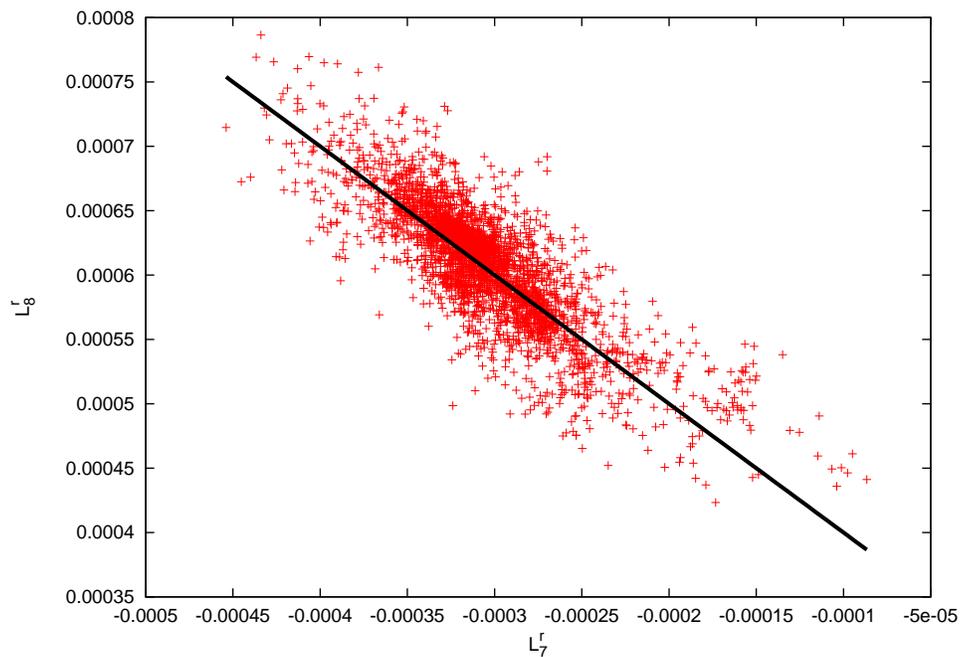}
\end{center}
\caption{\label{fig:L78}
The correlations between $L_7^r$ and $L_8^r$,
the curve shown is $L_8^r = -(L_7^r-0.0003)$.}
\end{figure}

There are also correlation between the fitted values of the $L_i^r$.
$L_1^r$, $L_2^r$ and $L_3^r$ show a reasonable correlation among themselves
but are essentially not correlated with the others.
There are weaker correlations between $L_4^r$ and $L_6^r$
and between $L_7^r$ and $L_8^r$. These correlations are shown in
Figures~\ref{fig:L12}, \ref{fig:L13}, \ref{fig:L23}, \ref{fig:L46},
and \ref{fig:L78}. We have shown a curve in all plots guiding the eye
as well and given it in the caption of the figure.

Note that throughout this section we have considered all fits 
with a $\chi^2<1$ to be essentially possible.

\section{Conclusions}
\label{conclusions}

In this paper we have shown the results for a new global fit of the $L^r_i$ at
NNLO, with techniques similar to the ones in
\cite{Amoros:2000mc,Amoros:2001cp}.
Different treatments of the $p^6$ coupling constants have been
considered: the resonance estimate of \cite{Amoros:2000mc}, the results of
\cite{Jiang:2009uf} and the use of randomly selected $C^r_i$. The results
are difficult to interpret and unexpected. 
All the fits that have been performed using the NNLO
couplings from \cite{Amoros:2000mc} or \cite{Jiang:2009uf} show both strong and weak
points. The fits obtained from the randomly selected $C^r_i$ are too
different from each other to draw a final conclusion, although they 
give a rough indication on where we can expect to find the values of $L^r_i$.
They also provide a proof of principle that with reasonable values of the $C_i^r$
a reasonably convergent series for $SU(3)$ ChPT can be obtained for
many quantities.

The fit that presents the least discrepancies and best convergence of the chiral
expansion is fit All in Table~\ref{tab:fitall}, which has been obtained with
the resonance estimate of the $C^r_i$. It succeeds in fitting many observables
like the $\pi\pi$ and $\pi K$ scattering parameters and the slope of the
scalar formfactor of the pion. It also reproduces quite well the experimental
results for the $f_s$ and $g_p$ $K_{\ell 4}$ formfactors although it does not
predict the curvature of $f_s$. The perturbative expansions for masses and
decay constants reported in (\ref{massconvfitAll}) and (\ref{decayconvfitAll})
look suspicious but are acceptable. 
On the other hand it does not satisfy the large $N_c$
relation $2L^r_1\approx L^r_2$ and it fails in well constraining the $L^r_4$
and $L^r_6$ values. Finally its prediction for $\bar{\ell_2}$ is
far from the current estimate of that constant. We have at present not included
many results from lattice QCD. We expect that this should improve in a few years
allowing for another step forward in confronting ChPT with data.

\section*{Acknowledgments}

We thank Juerg Gasser, Gerhard Ecker, Veronique Bernard, Emilie Passemar, Antonio Pich and
Gilberto Colangelo for discussions.
This work is supported in part by the European Community-Research
Infrastructure Integrating Activity ``Study of Strongly Interacting Matter'' 
(HadronPhysics2, Grant Agreement n. 227431)
and the Swedish Research Council grants 621-2008-4074 and 621-2010-3326.

\appendix

\section{$C_i$ values}
\label{app:Ci}

The aim of this section is to present the values used for the $C_i^r$. 
We only present results for the $C_i^r$ that actually contribute to the observables we have
included in the fits and we indicate with the superscript $*$ the ones that are subleading in $N_c$.
They can all be found in Table~\ref{tab:Ci}
The column labelled reso is the resonance exchange estimate of Section~\ref{sect:Cires}.
The column labelled CQM is the estimate of \cite{Jiang:2009uf} as discussed in Section~\ref{sect:chinaCi}.
The values directly from their model can be found in Table IV of \cite{Jiang:2009uf}. The values
in the column labelled CQM have been multiplied with nomalization factor $\alpha=0.27$
from the fit in the right column of Table~\ref{tab:chinaCi}. The normalization from the
fit to the linear NA48/2 input, Table~\ref{tab:chinaCilin} is a little smaller but essentially
the same. The last two columns are from the random walk/simulated annealing estimates for the $C_i^r$
reported in Section~\ref{sect:randomCi} from the best $\chi^2$ found starting from the resonance
estimate and a fully random starting point. These are the $C_i^r$ for the fits reported
in Table~\ref{tab:random}.

The main purpose is to show the typical sizes of the $C_i^r$ we obtained for the fits and that
the pattern for the good fits can be quite different.
\begin{table}
\begin{center}
\begin{tabular}{|l|cccc||l|cccc|}
\hline
 $i$& reso & CQM & best & best & $i$&reso    &    CQM    &  best & best\\
    &      &     & reso & rand &    &        &            & reso & rand\\
\hline
 $ 1 $ & $ 1.216$ & $ 0.866$ & $ 1.683$ & $ -0.733$&     $ 25 $ & $ -1.838$ & $ -1.366$ & $ -1.452$ & $ 1.282$\\    
 $ 2^* $ & $ 0.000$ & $ 0.000$ & $ 0.113$ & $ 0.280$&     $ 26 $ & $ -0.284$ & $ 0.765$ & $ -0.597$ & $ -0.485$\\    
 $ 3 $ & $ 0.000$ & $ -0.011$ & $ 0.324$ & $ 0.084$&     $ 27^* $ & $ -0.261$ & $ -0.352$ & $ -0.228$ & $ -0.155$\\  
 $ 4 $ & $ 1.452$ & $ 0.708$ & $ 2.225$ & $ 1.266$&      $ 28^* $ & $ 0.135$ & $ 0.069$ & $ 0.177$ & $ 0.147$\\      
 $ 5 $ & $ 0.619$ & $ -0.231$ & $ 0.779$ & $ 1.147$&     $ 29 $ & $ -1.363$ & $ -0.704$ & $ -1.907$ & $ -0.785$\\   
 $ 6^* $ & $ 0.000$ & $ 0.000$ & $ -0.307$ & $ -0.050$&   $ 30^* $ & $ 0.270$ & $ 0.137$ & $ 0.165$ & $ 0.545$\\      
 $ 7^* $ & $ 0.000$ & $ 0.000$ & $ 0.350$ & $ -0.003$&    $ 31 $ & $ -0.616$ & $ -0.144$ & $ -0.389$ & $ 1.310$\\    
 $ 8 $ & $ 0.619$ & $ 0.528$ & $ 1.434$ & $ 0.615$&      $ 32^* $ & $ -0.002$ & $ 0.041$ & $ 0.291$ & $ 0.356$\\     
 $ 9^* $ & $ 0.000$ & $ 0.000$ & $ 0.148$ & $ -0.194$&    $ 33^* $ & $ 0.208$ & $ 0.021$ & $ 0.291$ & $ -0.102$\\     
 $ 10 $ & $ -1.239$ & $ -0.240$ & $ -0.164$ & $ 0.307$&  $ 34 $ & $ 1.432$ & $ 0.363$ & $ 2.321$ & $ 1.077$\\       
 $ 11^* $ & $ 0.000$ & $ 0.000$ & $ -0.112$ & $ -0.344$&  $ 35^* $ & $ -0.009$ & $ 0.039$ & $ 0.236$ & $ 0.146$\\     
 $ 12 $ & $ -0.619$ & $ -0.078$ & $ -1.358$ & $ -0.512$& $ 36^* $ & $ 0.000$ & $ 0.000$ & $ -0.169$ & $ -0.087$\\    
 $ 13^* $ & $ 0.000$ & $ 0.000$ & $ 0.265$ & $ -0.002$&   $ 63 $ & $ 0.619$ & $ 0.683$ & $ 0.665$ & $ 0.776$\\       
 $ 14 $ & $ 0.000$ & $ -0.190$ & $ -0.759$ & $ -0.828$&  $ 64^* $ & $ 0.000$ & $ 0.000$ & $ 0.389$ & $ -0.095$\\     
 $ 15^* $ & $ 0.000$ & $ 0.000$ & $ -0.228$ & $ -0.233$&  $ 65 $ & $ 1.239$ & $ -0.555$ & $ 0.611$ & $ 0.432$\\      
 $ 16^* $ & $ 0.000$ & $ 0.000$ & $ 0.007$ & $ 0.063$&    $ 66 $ & $ 1.049$ & $ 0.391$ & $ 1.703$ & $ 0.416$\\       
 $ 17 $ & $ 0.000$ & $ 0.002$ & $ 0.125$ & $ 1.121$&     $ 67^* $ & $ 0.000$ & $ 0.000$ & $ -0.304$ & $ 0.017$\\     
 $ 18^* $ & $ -0.202$ & $ -0.128$ & $ -0.284$ & $ -0.063$&$ 68^* $ & $ 0.000$ & $ 0.000$ & $ 0.002$ & $ 0.536$\\      
 $ 19 $ & $ 0.001$ & $ -0.110$ & $ -0.411$ & $ -1.147$&  $ 69 $ & $ -0.577$ & $ -0.196$ & $ -0.664$ & $ -0.784$\\   
 $ 20^* $ & $ -0.002$ & $ 0.041$ & $ -0.335$ & $ -0.043$& $ 83 $ & $ 0.163$ & $ 0.016$ & $ -0.294$ & $ -0.553$\\     
 $ 21^* $ & $ 0.001$ & $ -0.014$ & $ 0.018$ & $ -0.088$&  $ 84^* $ & $ 0.000$ & $ 0.000$ & $ 0.357$ & $ -0.264$\\     
 $ 22 $ & $ -0.297$ & $ 0.062$ & $ 0.545$ & $ 1.117$&    $ 88 $ & $ -1.383$ & $ -1.249$ & $ -0.912$ & $ -0.331$\\   
 $ 23^* $ & $ 0.000$ & $ 0.000$ & $ 0.048$ & $ 0.269$&    $ 90 $ & $ 5.069$ & $ 0.557$ & $ 5.238$ & $ -0.204$\\      
 $ 24^* $ & $ 0.811$ & $ 0.370$ & $ 0.694$ & $ 0.013$&    &    &  &  &\\
\hline
\end{tabular}
\caption{\label{tab:Ci} The $C^r_i$ constants as obtained by several
  methods. In the table we quote $10^5\times C^r_i$ were the $i$ is given in
  the first column. The $i^*$ indicates a $N_c$ suppressed $C^r_i$. 
  The second column corresponds to the resonance model of
  Section~\ref{sect:Cires}. In the third one we quote the $C^r_i$ of \cite{Jiang:2009uf}
  multiplied by $\alpha=0.27$. The particular value of
  $\alpha$ is chosen fitting it to the input observables together with the
  $L^r_i$, as explained in Section~\ref{sect:chinaCi}. In the last two columns
  we quote two sets of $C^r_i$ corresponding to the best fits of the $L^r_i$
  when the $C^r_i$ are realesed, as explained in
  Section~\ref{sect:randomCi}. The column labeled best reso comes from initial
  values of the $C^r_i$ equal to the ones of the resonance estimate, while in
  the column labeled best rand  even the initial values of the $C^r_i$ are
  chosen randomly. The scale of
  renormalization for all the sets is $\mu=0.77$ GeV.}
\end{center}
\end{table}

\end{document}